\documentclass[10pt,journal]{IEEEtran}

\usepackage{array, calc, color, multicol}
\usepackage{amsmath}
\usepackage{cases}

\usepackage{algorithm}
\usepackage[noend]{algorithmic} 
\usepackage{graphics, epstopdf, graphicx, epsfig, psfrag, epsf, subfigure}
\usepackage{amssymb, amsfonts, amsmath, times}
\usepackage{relsize}
\usepackage{multicol,balance, verbatim}
\usepackage{textcomp, mathrsfs, stmaryrd, setspace}
\usepackage{amssymb}
\usepackage{wrapfig}

\newcommand{\ie}{{\em i.e., }}
\newcommand{\Ie}{{\em I.e., }}
\newcommand{\eg}{{\em e.g., }}
\newcommand{\Eg}{{\em E.g., }}

\newtheorem{theorem}{Theorem}
\newtheorem{lemma}[theorem]{Lemma}

\newtheorem{example}{Example}

\newcommand{\CCP}{{\tt C3P}}

\DeclareGraphicsExtensions{.eps, .pdf, .png, .jpg}   

\newcommand{\Nset}{\mathcal{N}}

\newcommand{\Rset}{\mathcal{R}}

\begin{document}
\title{Dynamic Heterogeneity-Aware Coded Cooperative Computation at the Edge}

\author{Yasaman~Keshtkarjahromi,~\IEEEmembership{Member,~IEEE,}
        Yuxuan~Xing,~\IEEEmembership{Student~Member,~IEEE,}        
        and~Hulya~Seferoglu,~\IEEEmembership{Member,~IEEE}
\thanks{This work was supported by the NSF under Grant CNS-1801708, the ARL under Grants W911NF-1820181 and W911NF-1710032, and the NIST under Grant 70NANB17H188. The preliminary results of this paper were presented in part at the IEEE International Conference on Network Protocols (ICNP), Cambridge, UK, Sep. 2018.}
\thanks{Y. Keshtkarjahromi was with the Department of Electrical and Computer Engineering, University of Illinois at Chicago. She is now an ORAU Postdoc Fellow. E-mail: y.keshtkar@gmail.com.} 
\thanks{Y. Xing, and H. Seferoglu are with the Department of Electrical and Computer Engineering, University of Illinois at Chicago, Chicago, IL, 60607, hulya@uic.edu, yxing7@uic.edu.}
}

\maketitle

\begin{abstract}
Cooperative computation is a promising approach for localized data processing at the edge, \eg for Internet of Things (IoT). Cooperative computation advocates that computationally intensive tasks in a device could be divided into sub-tasks, and offloaded to other devices or servers in close proximity. However, exploiting the potential of cooperative computation is challenging mainly due to the heterogeneous and time-varying nature of edge devices. Coded computation, which advocates mixing data in sub-tasks by employing erasure codes and offloading these sub-tasks to other devices for computation, is recently gaining interest, thanks to its higher reliability, smaller delay, and lower communication costs. In this paper, we develop a coded cooperative computation framework, which we name Coded Cooperative Computation Protocol (\CCP), by taking into account the heterogeneous resources of edge devices. \CCP \ dynamically offloads coded sub-tasks to helpers and is adaptive to time-varying resources. We show that (i) task completion delay of \CCP \ is very close to optimal coded cooperative computation solutions, (ii) the efficiency of \CCP \ in terms of resource utilization is higher than $99\%$, and (iii) \CCP \ improves task completion delay significantly as compared to baselines via both simulations and in a testbed consisting of real Android-based smartphones.
\end{abstract}

\section{\label{sec:introduction}Introduction}
Data processing is crucial for many applications at the edge including Internet of Things (IoT), but it could be computationally intensive and not doable if devices operate individually. One of the promising solutions to handle computationally intensive tasks is computation offloading, which advocates offloading tasks to remote servers or cloud. Yet, offloading tasks to remote servers or cloud could be luxury that cannot be afforded by most of the edge applications, where connectivity to remote servers can be lost or compromised, which makes localized processing crucial.

\begin{figure}[t!]
\begin{center}
\subfigure[Offloading sub-tasks from a collector to helpers]{ \scalebox{.24}{\includegraphics{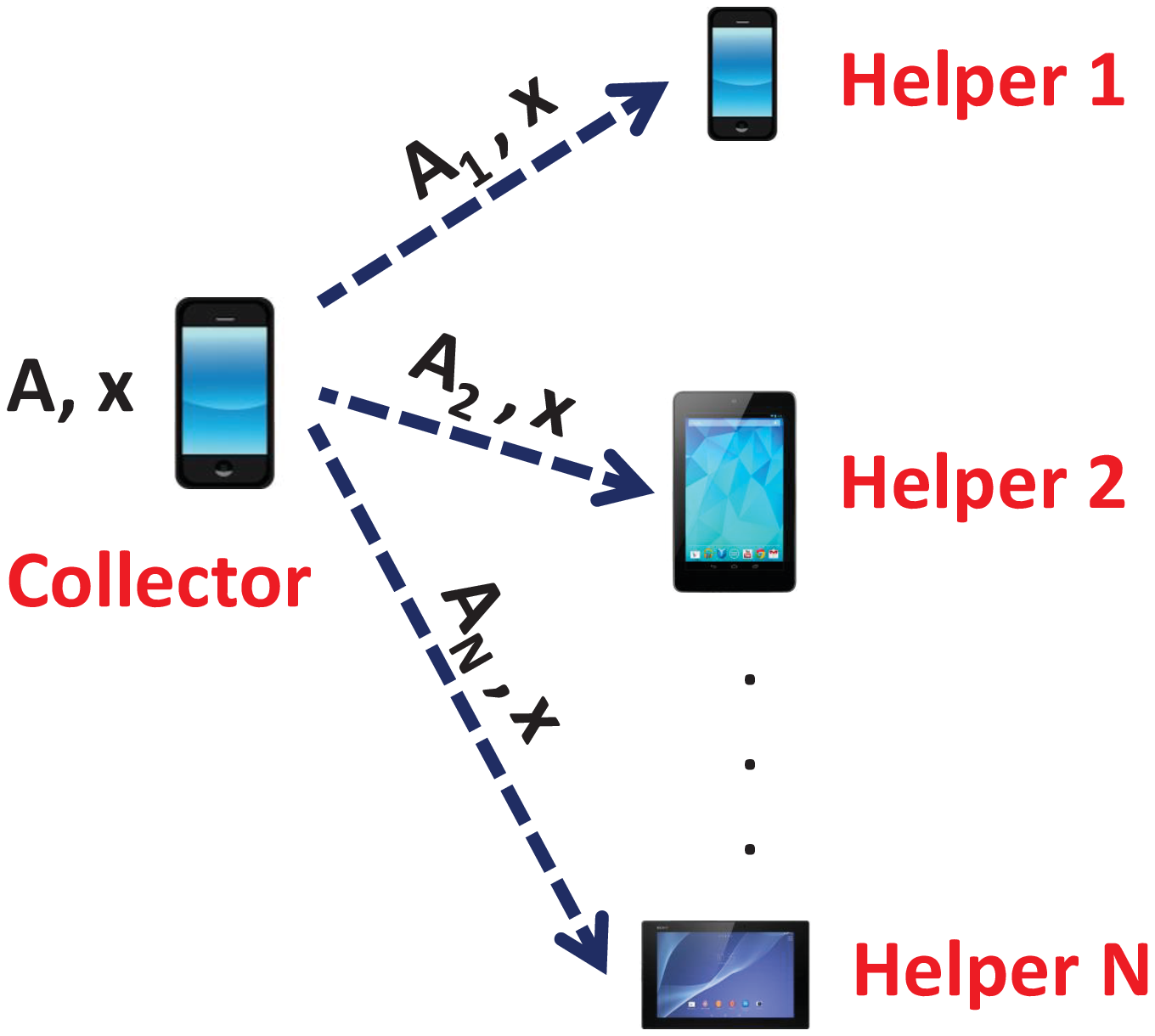}} } \hspace{10pt}
\subfigure[Helpers send computed sub-tasks back to the collector]{ \scalebox{.24}{\includegraphics{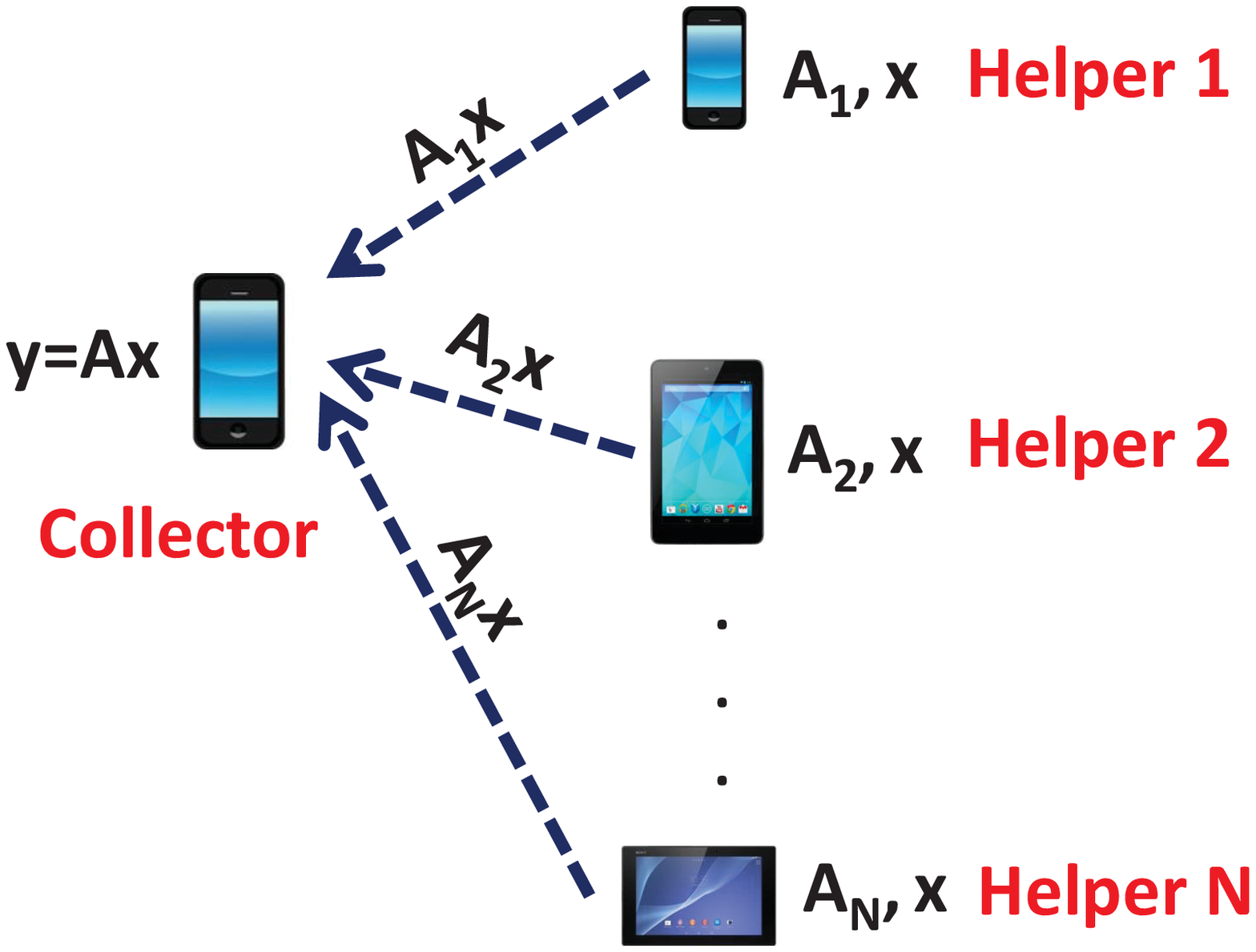}} }
\end{center}
\begin{center}
\vspace{-10pt}
\caption{\label{fig:uncoded} Cooperative computation to compute $\mathbf{y} = A\mathbf{x}$. (a) Matrix $A$ is divided into sub-matrices $A_1, A_2, ..., A_N$. Each sub-matrix along with the vector $\mathbf{x}$ is transmitted from the collector to one of the helpers. (b) Each helper computes the multiplication of its received sub-matrix with vector $\mathbf{x}$ and sends the computed value back to the collector.}
\end{center}
\end{figure}

Cooperative computation is a promising approach for  edge computing, where computationally intensive tasks in a device (collector device) could be offloaded to other devices (helpers) in close proximity as illustrated in Fig. \ref{fig:uncoded}.

These devices could be other IoT or mobile devices, local servers, or fog at the edge of the network \cite{MobileCloudlets}, \cite{AdHocCloudlets}.
However, exploiting the potential of cooperative computation is challenging mainly due to the heterogeneous and time-varying nature of the devices at the edge. Indeed, these devices may have different and time-varying computing power and energy resources, and could be mobile. Thus, our goal is to develop a dynamic, adaptive, and heterogeneity-aware cooperative computation framework by taking into account the heterogeneity and time-varying nature of devices at the edge.

We focus on the computation of linear functions. In particular, we assume that the collector's data is represented by a large matrix $A$ and it wishes to compute the product $\mathbf{y}=A \mathbf{x}$, for a given vector $\mathbf{x}$, Fig. \ref{fig:uncoded}. In fact, matrix multiplication forms the atomic function computed over many iterations of several signal processing, machine learning, and optimization algorithms, such as gradient descent based algorithms, classification algorithms, etc. \cite{SpeedingUp, ref_51, ref_52, ref_53}. 

In cooperative computation setup, matrix $A$ is divided into sub-matrices $A_1,A_2,...,A_N$ and each sub-matrix along with the vector $\mathbf{x}$ is transmitted from the collector to one of the helpers, Fig. \ref{fig:uncoded}(a). Helper $n$ computes $A_n \mathbf{x}$, and transmits the computed result back to the collector, Fig. \ref{fig:uncoded}(b), who can process all returned computations to obtain the result of its original task; \ie the calculation of $\mathbf{y}=A\mathbf{x}$.

Coding in computation systems is recently gaining interest in large scale computing environments, and it advocates higher reliability and smaller delay \cite{SpeedingUp}. In particular, coded computation (\eg by employing erasure codes) mixes data in sub-tasks and offloads these coded sub-tasks for computation, which improves delay and reliability. The following canonical example inspired from \cite{SpeedingUp} demonstrates the effectiveness of coded computation.

\begin{example} \label{ex:ex1}
Let us consider that a collector device would like to calculate $\mathbf{y}=A\mathbf{x}$ with the help of three helper devices (helper $1$, helper $2$, and helper $3$), where the number of rows in $A$ is $6$. Let us assume that each helper has a different runtime; helper $1$ computes each row in $1$ unit time, while the second and the third helpers require $2$ and $10$ units of time for computing one row, respectively. Assuming that these runtimes are random and not known a priori, one may divide $A$ to three sub-matrices; $A_1$, $A_2$, and $A_3$; each with $2$ rows. Thus, the completion time of these sub-matrices becomes $2$, $4$, and $20$ at helpers $1$, $2$, and $3$, respectively. Since the collector should receive all the calculated sub-matrices to compute its original task; \ie $\mathbf{y}=A\mathbf{x}$, the total task completion delay becomes $\max(2,4,20)=20$.

As seen, helper $3$ becomes a bottleneck in this scenario, which can be addressed using coding. In particular, $A$ could be divided into two sub-matrices $A_1$ and $A_2$; each with $3$ rows. Then, $A_1$ and $A_2$ could be offloaded to helpers $1$ and $2$, and $A_1 + A_2$ could be offloaded to helper $3$. In this setup, runtimes become $3$, $6$, and $30$ at helpers $1$, $2$, and $3$, respectively. However, since the collector requires reply from only two helpers to compute $\mathbf{y}=A\mathbf{x}$ thanks to coding, the total task completion delay becomes $\max(3,6)=6$. As seen, the task completion delay reduces to $6$ from $20$ with the help of coding.
\hfill $\Box$
\end{example}

The above example demonstrates the benefit of coding for cooperative computation. However, offloading sub-tasks with equal sizes to all helpers, without considering their heterogeneous resources is inefficient. Let us consider the same setup in Example~\ref{ex:ex1}. If $A_1$ with $4$ rows and $A_2$ with $2$ rows are offloaded to helper $1$ and helper $2$, respectively, and helper $3$ is not used, the task completion delay becomes $\max(4,4)=4$, which is the smallest possible delay in this example. Furthermore, the resources of helper $3$ are not wasted, which is another advantage of taking into account the heterogeneity as compared to the above example. As seen, it is crucial to divide and offload matrix $A$ to helpers by taking into account the heterogeneity of resources.

Indeed, a code design mechanism under such a heterogeneous setup is developed in \cite{HCMM}, where matrix $A$ is divided, coded, and offloaded to helpers by taking into account heterogeneity of resources. However, available resources at helpers are generally not known by the collector a priori and may vary over time, which is not taken into account in \cite{HCMM}. For example, the runtime of helper $1$ in Example~\ref{ex:ex1} may increase from $1$ to $20$ while computing (\eg it may start running another computationally intensive task), which would increase the total task completion delay. Thus, it is crucial to design a coded cooperation framework, which is dynamic and adaptive to heterogeneous and time-varying resources, which is the goal of this paper.

In this paper, we design a coded cooperative computation framework for edge computing. In particular, we design a Coded Cooperative Computation Protocol (\CCP), which packetizes rows of matrix $A$ into packets, codes these packets using Fountain codes, and determines how many coded packets each helper should compute dynamically over time. We provide theoretical analysis of \CCP 's task completion delay and efficiency, and evaluate its performance via simulations as well as in a testbed consisting of real Android-based smartphones as compared to baselines. The following are the key contributions of this work:
\begin{itemize}
\item We formulate the coded cooperative computation problem as an optimization problem. We investigate the non-ergodic and static solutions of this problem. As a dynamic solution to the optimization problem, we develop a coded cooperative computation protocol (\CCP), which is based on Automatic Repeat reQuest (ARQ) mechanism. In particular, a collector device offloads coded sub-tasks to helpers gradually, and receives Acknowledgment (ACK) after each sub-task is computed. Depending on the time difference between offloading a sub-task to a helper and its ACK, the collector estimates the runtime of the helpers, and offloads more/less tasks accordingly. This makes \CCP \ dynamic and adaptive to heterogeneous and time-varying resources at helpers.
\item We characterize the performance of \CCP \ as compared to the non-ergodic and static solutions, and show that (i) the gap between the task completion delays of \CCP \ and the non-ergodic solution is finite even for large number of sub-tasks, \ie $R \to \infty$
    , and (ii) the task completion delay of \CCP \ is approximately equal to the static solution for large numbers of sub-tasks. We also analyze the efficiency of \CCP \ in each helper in closed form, where the efficiency metric represents the effective utilization of resources at each helper.
\item We evaluate \CCP \ via simulations as well as in a testbed consisting of real Android-based smartphones and show that (i) \CCP \ improves task completion delay significantly as compared to baselines, and (ii) the efficiency of \CCP \ in terms of resource utilization is higher than $99\%$.

\end{itemize}

The structure of the rest of this paper is as follows. Section \ref{sec:System Model} presents the coded cooperative computation problem formulation. Section \ref{sec:ProbSol} presents the ergodic and static solutions to coded cooperative computation problem and the design of \CCP. Section \ref{sec:CPP_algorithm} provides the performance analysis of \CCP.
Section \ref{sec:Simulation} presents the performance evaluation of \CCP.
Section \ref{sec:LitratureReview} presents related work.
Section \ref{sec:conclusion} concludes the paper. 

\section{\label{sec:System Model} Problem Formulation}
{\em Setup.} We consider a setup shown in Fig.~\ref{fig:uncoded}, where the collector device offloads its task to helpers in the set $\Nset$ (where $N = |\Nset|$) via device-to-device (D2D) links such as Wi-Fi Direct and/or Bluetooth. In this setup, all devices could potentially be mobile, so the encounter time of the collector with helpers varies over time. \Ie the collector can connect to less than $N$ helpers at a time.

{\em Application.} As we described in Section~\ref{sec:introduction}, we focus on computation of linear functions; \ie the collector wishes to compute $\mathbf{y}=A\mathbf{x}$ where $A = (a_{i,j}) \in \mathbb{R}^{R \times R}$, and $\mathbf{x} = (x_{i,j}) \in \mathbb{R}^{R \times 1}$. Our goal is to determine sub-matrix $A_n = (a_{i,j}) \in \mathbb{R}^{r_n \times R}$ that will be offloaded to helper $n$, where $r_n$ is an integer. 

{\em Coding Approach.} We use Fountain codes \cite{LTCodes}, \cite{RaptorCodes}, which are ideal in our dynamic coded cooperation framework thanks to their rateless property, low encoding and decoding complexity, and low overhead. In particular, the encoding and decoding complexity of Fountain codes could be as low as $O(R\log(R))$ for LT codes and $O(R)$ for Raptor codes and the coding overhead could be as low as $5\%$ \cite{Fountain}. We note that Fountain codes perform better than (i) repetition codes thanks to randomization of sub-tasks by mixing them, (ii) maximum distance separable (MDS) codes as MDS codes require a priori task allocation (due to their block coding nature) and are not suitable for the dynamic and adaptive framework that we would like to develop, and (iii) network coding as the decoding complexity of network coding is too high \cite{RLNC}, which introduces too much computation overhead at the collector which obsoletes the computation offloading benefit.

{\em Packetization.} In particular, we packetize each row of $A$ into a packet and create $R$ packets; $\Gamma = \{ \rho_1, \rho_2, \ldots, \rho_R\}$. These packets are used to create Fountain coded packets, where $\nu_i$ is the $i$th coded packet. The coded packet $\nu_i$ is transmitted to a helper, where the helper computes the multiplication of $\nu_i \mathbf{x}$ and sends the result back to the collector. $R+K$ coded computed packets are required at the collector to decode the coded packets, where $K$ is the coding overhead. Let $p_{n,i}$ be the $j$th coded packet generated by the collector and the $i$th coded packet transmitted to helper $n$; $p_{n,i}=\nu_j, j \ge i$.

{\em Delay Model.} Each transmitted packet $p_{n,i}$ experiences transmission delay between the collector and helper $n$ as well as computing delay at helper $n$. Also, the computed packet $p_{n,i} \mathbf{x}$ experiences transmission delay while transmitted from helper $n$ to the collector. The average round trip time (RTT) of sending a packet to helper $n$ and receiving the computed packet, is characterized as $RTT_n^{\text{data}}$. The runtime of packet $p_{n,i}$ at helper $n$ is a random variable denoted by $\beta_{n,i}$.\footnote{Our framework is compatible with any delay distribution, but for the sake of characterizing the efficiency of our algorithm, and simulating its task completion delay, we use shifted exponential distribution in Sections \ref{sec:efficiency} and  \ref{sec:Simulation}.} Assuming that $r_n$ packets are offloaded to helper $n$, the total task completion delay for helper $n$ to receive $r_n$ coded packets, compute them, and send the results back to the collector becomes $D_n$, which is expressed as $D_n = RTT_n^{\text{data}} + \sum_{i=1}^{r_n} \beta_{n,i}$. Note that $RTT_n^{\text{data}}$ in this formulation is due to transmitting the first packet $p_{n,1}$ and receiving the last computed packet $p_{n,r_n}\mathbf{x}$. The other packets can be transmitted while helpers are busy with processing packets; it is why we do not sum $RTT_n^{\text{data}}$ across packets.

{\em Problem Formulation.} Our goal is to determine the task offloading set $\Rset = \{r_1, \ldots, r_N\}$ 
that minimizes the total task completion delay, \ie we would like to dynamically determine $\Rset$ that solves the following optimization problem:
\begin{align} \label{eq:opt_hs}
\min_{\Rset} & \max_{n \in \Nset} D_n \nonumber \\
\mbox{subject to  } & \sum_{n=1}^{N} r_n = R,  r_n \in \mathbb{N}, \forall n \in \Nset.
\end{align}
The objective of the optimization problem in (\ref{eq:opt_hs}) is to minimize the maximum of per helper task completion delays, which is equal to $\max_{n \in \Nset} D_n$, as helpers compute their tasks in parallel. The constraint in (\ref{eq:opt_hs}) is a task conservation constraint that guarantees that resources of helpers are not wasted, \ie the sum of the received computed tasks from all helpers is equal to the number of rows of matrix $A$. Note that this constraint is possible thanks to coding.\footnote{We note that the optimal computation offloading problem, when coding is not employed, is formulated as $\min_{\Gamma_n}  \max_{n \in \Nset} (RTT_n^{\text{data}} + \sum_{i=1}^{|\Gamma_n|} \beta_{n,i})$  subject to $\cup_{n=1}^{N} \Gamma_n = \Gamma$ where $\Gamma_n\subset\Gamma$ is the set of packets offloaded to helper $n$. As seen, the optimization problem in (\ref{eq:opt_hs}) is more tractable as compared to this problem thanks to employing Fountain codes. 
} As we mentioned earlier, $R+K$ coded computed packets are required at the collector to decode the coded packets when we use Fountain codes. The constraint in (\ref{eq:opt_hs}) guarantees this requirement in an idealized scenario assuming that $K = 0$. The constraint $r_n \in \mathbb{N}$ makes sure that the number of tasks $r_n$ is an integer. The solution of (\ref{eq:opt_hs}) is challenging as (i) $D_n = RTT_n^{\text{data}} + \sum_{i=1}^{r_n} \beta_{n,i}$ is a random variable and not known a priori, and (ii) it is an integer programming problem.

\section{\label{sec:ProbSol} Problem Solution \& \CCP \ Design}

In this section, we investigate the solution of (\ref{eq:opt_hs}) for non-ergodic, static, and dynamic setups.

\subsection{\label{sec:nonErgodic} Non-Ergodic Solution}
Let us assume that the solution of (\ref{eq:opt_hs}) is
\begin{align}\label{eq:non_ergodic_solution}
T^\text{best}=\max_{n \in \Nset} \Big(RTT_n^{\text{data}}+ \sum_{i=1}^{r_n^{\text{best}}} \beta_{n,i}\Big),
\end{align} where $r_n^{\text{best}}$ $=$ $\operatorname*{argmin}_{r_n \in \mathbb{N}} \max_{n \in \Nset}$ $\Big(RTT_n^{\text{data}}+\sum_{i=1}^{r_n}$ $\beta_{n,i}\Big)$. We note that (\ref{eq:non_ergodic_solution}) is a non-ergodic solution as it requires the perfect knowledge of $\beta_{n,i}$ a priori. Although we do not have a compact solution of $T^\text{best}$,  the solution in (\ref{eq:non_ergodic_solution}) will behave as a performance benchmark for our dynamic and adaptive coded cooperative computation framework in Section \ref{sec:CCP_vs_NonErgodic}.

\subsection{\label{sec:static} Static Solution}
We assume that $RTT_n^{\text{data}}$ becomes negligible as compared to $\sum_{i=1}^{r_n} \beta_{n,i}$. This assumption holds in  practical scenarios with large $R$, and/or when transmission delay is smaller than processing delay. Then, $D_n$ can be approximated as $\sum_{i=1}^{r_n} \beta_{n,i}$, and the optimization problem in (\ref{eq:opt_hs}) becomes
\begin{align}\label{eq:D_n2}
 \min_{\Rset}  & \max_{n \in \Nset} \sum_{i=1}^{r_n} \beta_{n,i} \nonumber  \\
\mbox{subject to  } & \sum_{n=1}^{N} r_n = R, r_n \in \mathbb{N}, \forall n \in \Nset.
\end{align}

\begin{figure*}[h!]
\centering
\subfigure[
Ideal case]{ \scalebox{.28}{\includegraphics{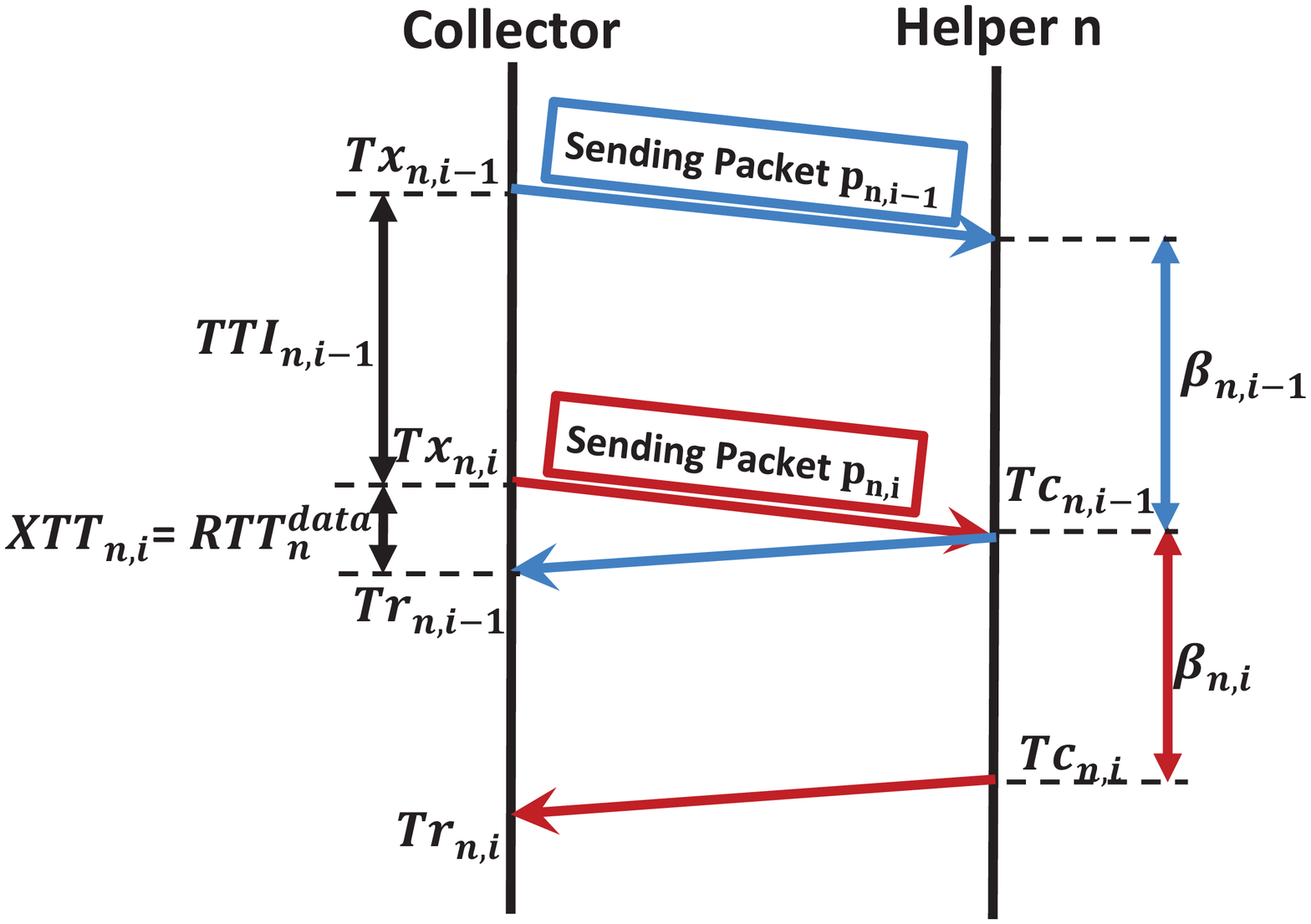}} }
\subfigure[
Underutilized case]{ \scalebox{.28}{\includegraphics{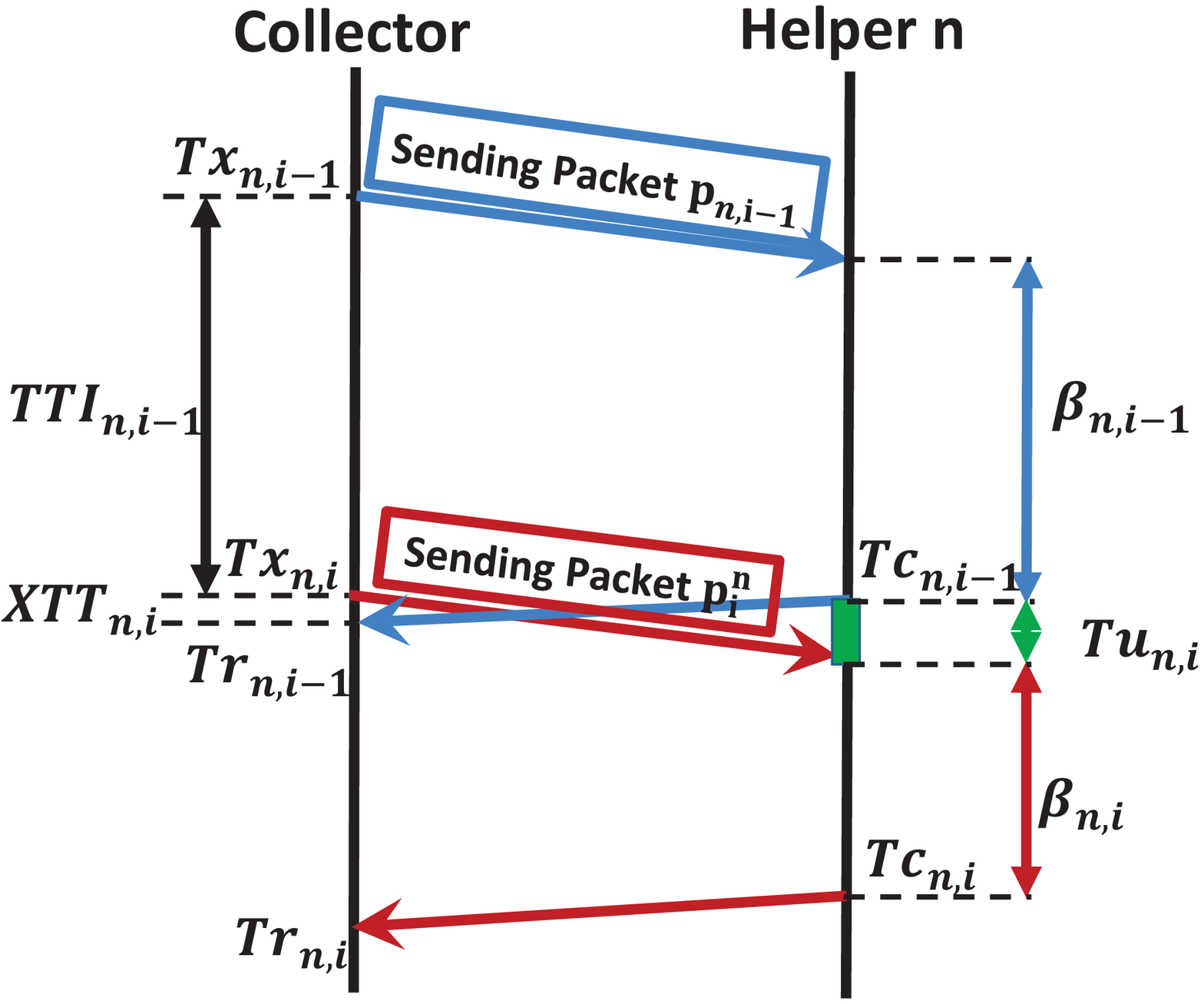}} }
\subfigure[
Congested case]{ \scalebox{.28}{\includegraphics{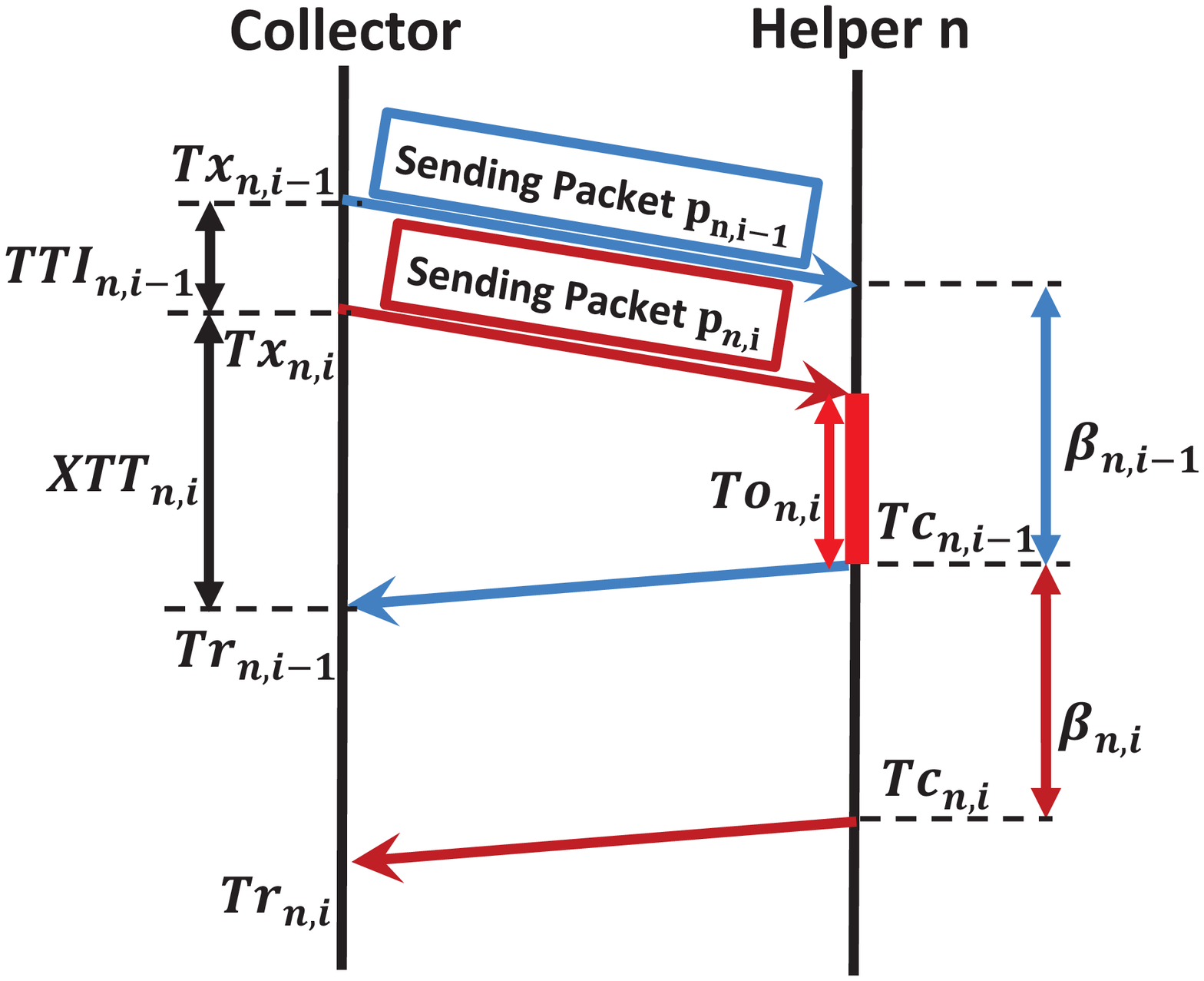}} }
\caption{
Different states of the system: (a) ideal case, 
(b) underutilized case, and (c) congested case.}
\label{fig:CPP_Scenarios}
\end{figure*}

As a static solution, we solve the expected value of the objective function in (\ref{eq:D_n2}) by relaxing the integer constraint, \ie $r_n \in \mathbb{N}$. The expected value of the objective function of (\ref{eq:D_n2}) is expressed as $E[\max_{n \in \Nset}$ $\sum_{i=1}^{r_n} $ $\beta_{n,i}]$, which is greater than or equal to $\max_{n \in \Nset}$ $\sum_{i=1}^{r_n} E[\beta_{n,i}] = \max_{n \in \Nset} r_n E[\beta_{n,i}]$ (noting that $\max(.)$ is a convex function, so $E[\max(.)] \ge \max(E[.])$), where expectation is across the packets. Assuming that the average task completion delay is $T = E[\max_{n \in \Nset} \sum_{i=1}^{r_n} \beta_{n,i}] \ge \max_{n \in \Nset}$ $r_n$ $E[\beta_{n,i}]$,  (\ref{eq:D_n2}) is converted to
\begin{align}\label{eq:D_n3}
 \min_{\Rset}  \quad & T  \nonumber  \\
 \mbox{subject to  } & r_n E[\beta_{n,i}] \leq T, \forall n\in\Nset \nonumber  \\
 & \sum_{n=1}^{N} r_n = R.
\end{align}
We solve (\ref{eq:D_n3}) using Lagrange relaxation (we omit the steps of the solution as it is straightforward); the optimal task offloading policy becomes
\begin{align}\label{eq:rnstar}
r_n^{\text{static}} = \frac{R}{E[\beta_{n,i}]\sum_{n=1}^{N} \frac{1}{E[\beta_{n,i}]}},
\end{align} and the optimal task completion delay becomes $T^{\text{static}} = \frac{R}{\sum_{n=1}^{N} \frac{1}{E[\beta_{n,i}]}}$. Although the solution in (\ref{eq:rnstar}) is an optimal solution of (\ref{eq:D_n3}), the algorithm that offloads $r_n^{\text{static}}$ sub-tasks to helper $n$ a priori (static allocation) loses optimality as it is not adaptive to the time-varying nature of resources (\ie $\beta_{n,i}$). Next, we introduce our Coded Cooperative Computation Protocol (\CCP) that is dynamic and adaptive to time-varying resources and approaches to the optimal solution in (\ref{eq:rnstar}) with increasing $R$.

\subsection{\label{sec:CPP_algorithm} Dynamic Solution: \CCP}

We consider the system setup in Fig.~\ref{fig:uncoded}, where the collector connects to $N$ helpers.
In this setup, the collector device offloads coded packets gradually to helpers, and receives two ACKs for each packet; one confirming the receipt of the packet by the helper, and the second one (piggybacked to the computed packet $p_{n,i} \mathbf{x}$) showing that the packet is computed by the helper. Inspired by ARQ mechanisms \cite{ARQ}, the collector transmits more/less coded packets based on the frequency of the received ACKs.

In particular, we define the transmission time interval $TTI_{n,i}$ as the time interval between sending two consecutive packets, $p_{n,i}$ and $p_{n,i+1}$, to helper $n$ by the collector. The goal of our mechanism is to determine the best $TTI_{n,i}$ that reduces the task completion delay and increases helper efficiency (\ie exploiting the full potential of the helpers while not overloading them). 

{\em $TTI_{n,i}$ in an ideal scenario.} Let $Tx_{n,i}$ be the time that $p_{n,i}$ is transmitted from the collector to helper $n$, $Tc_{n,i}$ be the time that helper $n$ finishes computing $p_{n,i}$, and $Tr_{n,i}$ be the time that the computed packet (\ie by abusing the notation $p_{n,i} \mathbf{x}$) is received by the collector from helper $n$. We assume that the time of transmitting the first packet to each helper, \ie $p_{n,1}, \forall n \in \Nset$, is zero; \ie $Tx_{n,1} = 0$, $\forall n \in \Nset$.

Let us first consider the ideal scenario, Fig. \ref{fig:CPP_Scenarios}(a), where $TTI_{n,i}$ is equal to $\beta_{n,i}$ for all packets that are transmitted to helper $n$. Indeed, if $TTI_{n,i} > \beta_{n,i}$, Fig. \ref{fig:CPP_Scenarios}(b), helper $n$ stays idle, which reduces the efficient utilization of resources and increases the task completion delay. On the other hand, if $TTI_{n,i} < \beta_{n,i}$, Fig. \ref{fig:CPP_Scenarios}(c), packets are queued at helper $n$. This congested (overloaded) scenario is not ideal, because the collector can receive enough number of packets before all queued packets in helpers are processed, which wastes resources.

{\em Determining $TTI_{n,i}$ in practice.} Now that we know that $TTI_{n,i} = \beta_{n,i}$ should be satisfied for the best system efficiency and smallest task completion delay, the collector can set $TTI_{n,i}$ to $\beta_{n,i}$. However, the collector does not know $\beta_{n,i}$ a priori as it is the computation runtime of packet $p_{n,i}$ at helper $n$. Thus, we should determine $TTI_{n,i}$ without explicit knowledge of $\beta_{n,i}$.

Our approach in \CCP \ is to estimate $\beta_{n,i}$ as $E[\beta_{n,i}]$, where expectation is taken over packets. We will explain how to calculate $E[\beta_{n,i}]$ later in this section, but before that let us explain how to use estimated $E[\beta_{n,i}]$ for setting $TTI_{n,i}$. It is obvious that if the computed packet $p_{n,i}\mathbf{x}$ is received at the collector before packet $p_{n,i+1}$ is transmitted from the collector to helper $n$, the helper will be idle until it receives packet $p_{n,i+1}$. Therefore, to better utilize resources at helper $n$, the collector should offload a new packet before or immediately after receiving the computed value of the previous packet, \ie $TTI_{n,i} \leq Tr_{n,i}-Tx_{n,i}$ should be satisfied as in Fig. \ref{fig:CPP_Scenarios}. Therefore, if the calculated $E[\beta_{n,i}]$ is larger than $Tr_{n,i} - Tx_{n,i}$, then we set $TTI_{n,i}$ as $Tr_{n,i} - Tx_{n,i}$ to satisfy this condition. In other words, $TTI_{n,i}$ is set to 
\begin{align} \label{eq:TTI_in}
TTI_{n,i} = \min ( Tr_{n,i} - Tx_{n,i}, E[ \beta_{n,i} ] ).
\end{align}

{\em Calculation of $E[\beta_{n,i}]$.} In \CCP, $E[\beta_{n,i}]$ is estimated using runtimes of previous packets:
\begin{align}\label{eq:Exp_beta}
  E[\beta_{n,i}] \approx \frac{\sum_{j=1}^{m_n} \beta_{n,i}}{m_n},
\end{align}
where $m_n$ is the number of computed packets received at the collector from helper $n$ before sending packet $p_{n,i+1}$. In order to calculate (\ref{eq:Exp_beta}), the collector device should have $\beta_{n,i}$ values from the previous offloaded packets. A straightforward approach would be putting timestamps on sub-tasks to directly access the runtimes $\beta_{n,i}$ at the collector. However, this approach introduces overhead on sub-tasks. Thus, we also developed a mechanism, where the collector device infers $\beta_{n,i}$ by taking into account transmission and ACK times of sub-tasks. The details of this approach is provided in Appendix A.

{\em \CCP \ in a nutshell.} The main goal of \CCP \ is to determine packet transmission intervals, $TTI_{n,i}$, according to (\ref{eq:TTI_in}), which is summarized in Algorithm~\ref{al:ccp}. Note that Algorithm~\ref{al:ccp} has also a timeout value defined in line \ref{al:Int_TO_3}, which is needed for unresponsive helpers. If helper $n$ is not responsive, $TTI_{n,i}$ is quickly increased as shown in line \ref{al:Int_TO_2} so that fewer and fewer packets could be offloaded to that helper. In particular, \CCP \ doubles $TTI_{n,i}$ when the timeout for receiving ACK occurs. This is inspired by additive increase multiplicative decrease strategy of TCP, where the number of transmitted packets are halved to backoff quickly when the system is not responding.

After $TTI_{n,i}$ is updated when a transmitted packet is ACKed or timeout occurs, this interval is used to determine the transmission times of the next coded packets. In particular, coded packets are generated and transmitted one by one to all helpers with intervals $TTI_{n,i}$ until (i) $TTI_{n,i}$ is updated with a new ACK packet or when timeout occurs, or (ii) the collector collects $R+K$ computed packets. Next, we characterize the performance of \CCP.

\begin{algorithm}[t!]
\caption{\CCP \ algorithm at the collector}
\label{al:ccp}
\begin{algorithmic}[1]
\STATE Initialize: $TO_n=\infty$, $\forall n \in \Nset$.

\WHILE {$R+K$ calculated packets have not been received}
 \IF {Calculated packet $p_{n,i} \mathbf{x}$ is received before timeout expires}
	\STATE Calculate $TTI_{n,i}$ according to (\ref{eq:Exp_beta}) and (\ref{eq:TTI_in}).
 \ELSE
	\STATE $TTI_{n,i}=2 \times TTI_{n,i}$. \label{al:Int_TO_2}	
 \ENDIF
 \STATE Update timeout as $TO_n = 2TTI_{n,i}$.  \label{al:Int_TO_3}	
\ENDWHILE
\end{algorithmic}
\end{algorithm}

\section{\label{sec:CPP_algorithm} Performance Analysis of \CCP}
\subsection{\label{sec:CCP_vs_NonErgodic}Performance of \CCP \ w.r.t. the Non-Ergodic Solution}

In this section, we analyze the gap between \CCP \ and the non-ergodic solution characterized in Section~\ref{sec:nonErgodic}. Let us first characterize the task completion delay of \CCP \ as
\begin{align}\label{eq:CCP_T}
  T^{\text{\CCP}} = \max_{n \in \Nset} \Big(RTT_n^{\text{data}}+\sum_{i=1}^{r_n^{\text{\CCP}}} (\beta_{n,i}+Tu_{n,i})\Big),
\end{align} where $r_n^{\text{\CCP}} = \operatorname*{argmin}_{r_n} \max_{n \in \Nset} \Big(RTT_n^{\text{data}}+\sum_{i=1}^{r_n} (\beta_{n,i}+Tu_{n,i})\Big)$, and $Tu_{n,i}$ is per packet under-utilization time at helper $n$, which occurs as \CCP \ does not have a priori knowledge of $\beta_{n,i}$, but it estimates $\beta_{n,i}$ and accordingly determines packet transmission times $TTI_{n,i}$ according to (\ref{eq:TTI_in}). The gap between $T^{\text{\CCP}}$ and $T^{\text{best}}$ in (\ref{eq:non_ergodic_solution}) is upper bounded by:
\begin{align} \label{eq:Gap_opt}
  T^{\text{\CCP}}-T^{\text{best}} & = \max_{n \in \Nset} \Big(RTT_n^{\text{data}}+\sum_{i=1}^{r_n^{\text{\CCP}}} (\beta_{n,i}+Tu_{n,i})\Big) \nonumber\\
  &-\max_{n \in \Nset} \Big(RTT_n^{\text{data}}+\sum_{i=1}^{r_n^{\text{best}}} \beta_{n,i}\Big) \nonumber\\
  & \leq \max_{n \in \Nset} \Big(RTT_n^{\text{data}}+\sum_{i=1}^{r_n^{\text{best}}} (\beta_{n,i}+Tu_{n,i})\Big) \nonumber \\
  &-\max_{n \in \Nset} \Big(RTT_n^{\text{data}}+\sum_{i=1}^{r_n^{\text{best}}} \beta_{n,i}\Big) \nonumber\\
  & \leq \max_{n \in \Nset} \Big(RTT_n^{\text{data}}+\sum_{i=1}^{r_n^{\text{best}}} \beta_{n,i}\Big)+\max_{n \in \Nset} \sum_{i=1}^{r_n^{\text{best}}}Tu_{n,i} \nonumber \\
  &-\max_{n \in \Nset} \Big(RTT_n^{\text{data}}+\sum_{i=1}^{r_n^{\text{best}}} \beta_{n,i}\Big) \nonumber\\
  & = \max_{n \in \Nset} \sum_{i=1}^{r_n^{\text{best}}}Tu_{n,i},
\end{align}
where the first inequality comes from $r_n^{\text{\CCP}}$ $=$ $\operatorname*{argmin}_{r_n}$ $\max_{n \in \Nset}$ $\Big(RTT_n^{\text{data}}+\sum_{i=1}^{r_n} (\beta_{n,i}+Tu_{n,i})\Big)$ and the second inequality comes from the fact that $\max(f(x)+g(x))\leq(\max(f(x))+\max(g(x)))$.\footnote{Note that in (\ref{eq:Gap_opt}), we assume that the runtime of packet $i$ at helper $n$ is the same in both the non-ergodic solution and \CCP, which is necessary for fair comparison.} As seen, the gap between \CCP \ and the non-ergodic solution is bounded with the sum of $Tu_{n,i}$. The next theorem characterizes $Tu_{n,i}$.

\begin{theorem} \label{theorem:CCP_vs_nonergodic}
$Tu_{n,i}$ is monotonically decreasing with increasing number of sub-tasks, and $\lim_{i\to\infty} Pr(Tu_{n,i}>0) \to 0 $.
\end{theorem}
{\em Proof:}
Let us first consider the following lemma that determines the conditions for having a positive $Tu_{n,i+1}$.
\begin{lemma} \label{th:underutilized_conditions}
The necessary and sufficient conditions to satisfy $Tu_{n,i+1}>0$ are
\begin{align}
    \sum_{j=i+1-k}^{i} \beta_{n,j} < kE[\beta_{n,i}], \forall k=1,2,\ldots,i
\end{align}
\end{lemma}
{\em Proof:}
The proof is provided in Appendix B. \hfill $\Box$

According to the conditions given in Lemma \ref{th:underutilized_conditions}, the probability of $Tu_{n,i}>0$ is calculated as:
\begin{align}\label{prob_underutilized_proof_1}
  Pr(Tu_{n,i}>0) = &\int_{0}^{E[\beta_{n,i}]}\int_{0}^{2E[\beta_{n,i}]-x_i} \ldots \int_{0}^{iE[\beta_{n,i}]-\sum_{j=2}^{i}\beta_{n,j}} \\ & f_{\beta_{n,1}, \ldots, \beta_{n,i}}(x_1, \ldots, x_i) dx_1 \ldots dx_i, \nonumber
\end{align}
where $f_{\beta_{n,1}, \ldots, \beta_{n,i}}(x_1, \ldots, x_i)$ is the joint probability density function of $(\beta_{n,1}, \ldots, \beta_{n,i})$. With the assumption that $\beta_{n,j}, j=1,2,...,i$ is from an i.i.d distribution, the joint probability distribution function of $\beta_{n,1}, \ldots, \beta_{n,i}$ is the product of $i$ probability distribution functions:
\begin{align}\label{prob_underutilized_proof}
  Pr(Tu_{n,i}>0) = &\int_{0}^{E[\beta_{n,i}]}\int_{0}^{2E[\beta_{n,i}]-x_i}...\int_{0}^{iE[\beta_{n,i}]-\sum_{j=2}^{i}x_j}\nonumber\\ &f_{\beta_{n,i}}(x_1)f_{\beta_{n,i}}(x_2)...f_{\beta_{n,i}}(x_i) dx_1 dx_2 ... dx_i\\
   = &  \int_{0}^{E[\beta_{n,i}]}f_{\beta_{n,i}}(x_i)\int_{0}^{2E[\beta_{n,i}]-x_i}f_{\beta_{n,i}}(x_{i-1})\nonumber\\
   &...\int_{0}^{(i-1)E[\beta_{n,i}]-\sum_{j=3}^i x_j} f_{\beta_{n,i}}(x_2) \nonumber \\
   &\int_{0}^{iE[\beta_{n,i}]-\sum_{j=2}^{i}x_j} f_{\beta_{n,i}}(x_1) dx_1 dx_2 ... dx_i\\
   < &\int_{0}^{E[\beta_{n,i}]}f_{\beta_{n,i}}(x_i)\int_{0}^{2E[\beta_{n,i}]-x_i} f_{\beta_{n,i}}(x_{i-1})\nonumber\\
   &...\int_{0}^{(i-1)E[\beta_{n,i}]-\sum_{j=3}^i x_j} f_{\beta_{n,i}}(x_2) dx_2 ... dx_i \\
   = &\int_{0}^{E[\beta_{n,i}]}f_{\beta_{n,i}}(x_{i-1})\nonumber\\
   &\int_{0}^{2E[\beta_{n,i}]-x_{i-1}}f_{\beta_{n,i}}(x_{i-2})...\nonumber\\
   &\int_{0}^{(i-1)E[\beta_{n,i}]-\sum_{j=2}^{i-1} x_j} f_{\beta_{n,i}}(x_1) dx_1 ... dx_{i-1}, \label{eq:PrTuni-1}
\end{align}
where the last inequality comes from the fact that $\int_{0}^{iE[\beta_{n,i}]-\sum_{j=2}^{i}x_j} f_{\beta_{n,i}}(x_1) dx_1$ is less than 1, because the probability density function is integrated over a finite range of variable $x_1$, and the last equality comes from a change of variables in the integrals. (\ref{eq:PrTuni-1}) is equal to $Pr(Tu_{n,i-1}>0)$ and thus $Pr(Tu_{n,i}>0) < Pr(Tu_{n,i-1}>0)$. Similarly, we can show that:
\begin{align}
  Pr(Tu_{n,j}>0) < Pr(Tu_{n,j-1}>0), \forall j=2,3,\ldots,i
\end{align}
From the above equation, we can conclude that as $i$ gets larger, $Pr(Tu_{n,i}>0)$ gets smaller, and $\lim_{i\to\infty} Pr(Tu_{n,i}>0) \to 0 $ is satisfied. This concludes the proof. \hfill $\Box$

We can conclude from Theorem~\ref{theorem:CCP_vs_nonergodic} that the rate of the increase in the gap between \CCP \ and the non-ergodic solution decreases with increasing the number of sub-tasks and eventually the rate becomes zero for $R \to \infty$. Therefore, the gap becomes finite even for $R \to \infty$.

\subsection{Performance of \CCP \ w.r.t the Static Solution}
In this section, we analyze the performance of \CCP \ as compared to the static solution characterized in Section~\ref{sec:static}. The next theorem characterizes the task completion delay of \CCP \ as well as the optimal task offloading policy.

\begin{theorem} \label{theorem:CCP_vs_static}
The task completion delay of \CCP \ approaches to
\begin{align}\label{eq:TCCP}
T^{\text{\CCP}} \approx \frac{R+K}{\sum_{n=1}^{N} \frac{1}{E[\beta_{n,i}]}},
\end{align} with increasing $R$ and the number of offloaded tasks to helper $n$ is approximated as
\begin{align}\label{eq:rnCCP}
r_n^{\text{\CCP}} \approx \frac{R+K}{E[\beta_{n,i}]\sum_{n=1}^{N} \frac{1}{E[\beta_{n,i}]}}.
\end{align}
\end{theorem}

{\em Proof:}  Proof is provided in Appendix C. \hfill $\Box$

Theorem~\ref{theorem:CCP_vs_static} shows that the task completion delay of \CCP \ is getting close to the static solution $T^{\text{static}}$ characterized in Section~\ref{sec:static} with increasing $R$. The gap between $T^{\text{static}}$ and $T^{\text{\CCP}}$ is $\frac{K}{\sum_{n=1}^{N} \frac{1}{E[\beta_{n,i}]}}$ which is due to the coding overhead of Fountain codes, which becomes negligible for large $R$.

\subsection{Performance of \CCP \ w.r.t. Repetition Codes}
In this section, we demonstrate the performance of \CCP \ as compared to repetition coding with Round-robin (RR) scheduling through an illustrative example. Repetition codes with RR scheduling works as follows. Uncoded packets from the set $\Gamma = \{ \rho_1, \rho_2, \ldots, \rho_R\}$ is offloaded to helpers one by one (in round robin manner) depending on their sequence in $\Gamma$. For example, $\rho_1$ is offloaded to helper 1, $\rho_2$ is offloaded to helper 2, and so on. When all the packets are offloaded from $\Gamma$, we start again from the first packet in the set (so it is a repetition coding). Note that whenever a packet is computed and a corresponding ACK is received, the packet is removed from $\Gamma$. Thus, this RR scheduling continues until $\Gamma$ becomes an empty set. We use $TTI_{n,i}$ in (\ref{eq:TTI_in}) to determine the next scheduling time for helper $n$. The next example demonstrates the benefit of \CCP \ as compared to this repetition coding mechanism with RR scheduling.

\begin{example} \label{ex:ex2}
We consider the same setup in Example~\ref{ex:ex1}. We assume that per-packet runtimes are $\beta_{1,1}=1, \beta_{1,2}=1, \beta_{1,3}=0.5, \beta_{1,4}=1, \beta_{1,5}=1.5$, $\beta_{2,1}=1.5, \beta_{2,2}=3.5$, and $\beta_{3,1}=3, \beta_{3,2}=2.5$, and the transmission times of packets are negligible.

\begin{figure*}[t!]
\begin{center}
\subfigure[Repetition codes with RR scheduling]{ \scalebox{.41}{\includegraphics{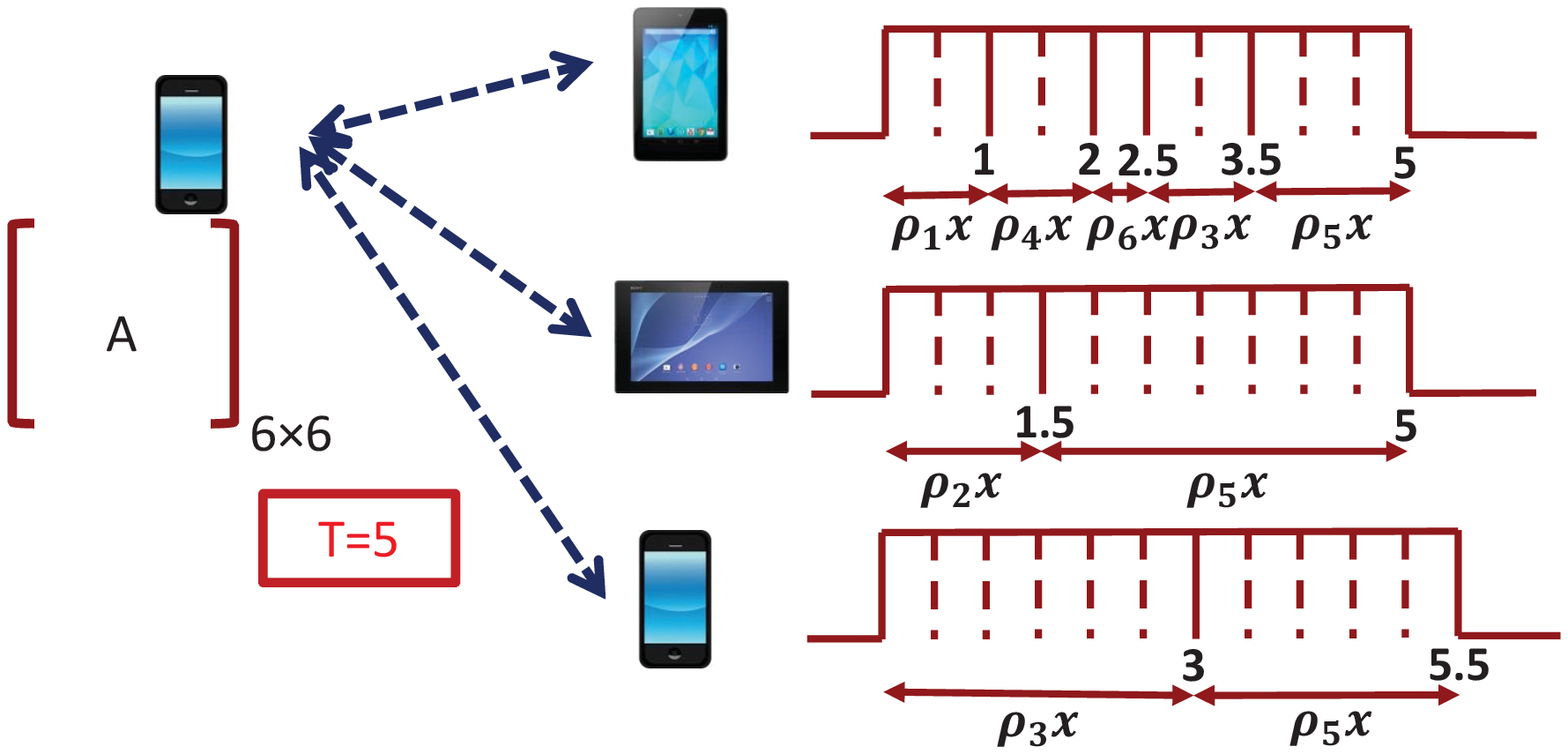}} }
\subfigure[\CCP]{ \scalebox{.41}{\includegraphics{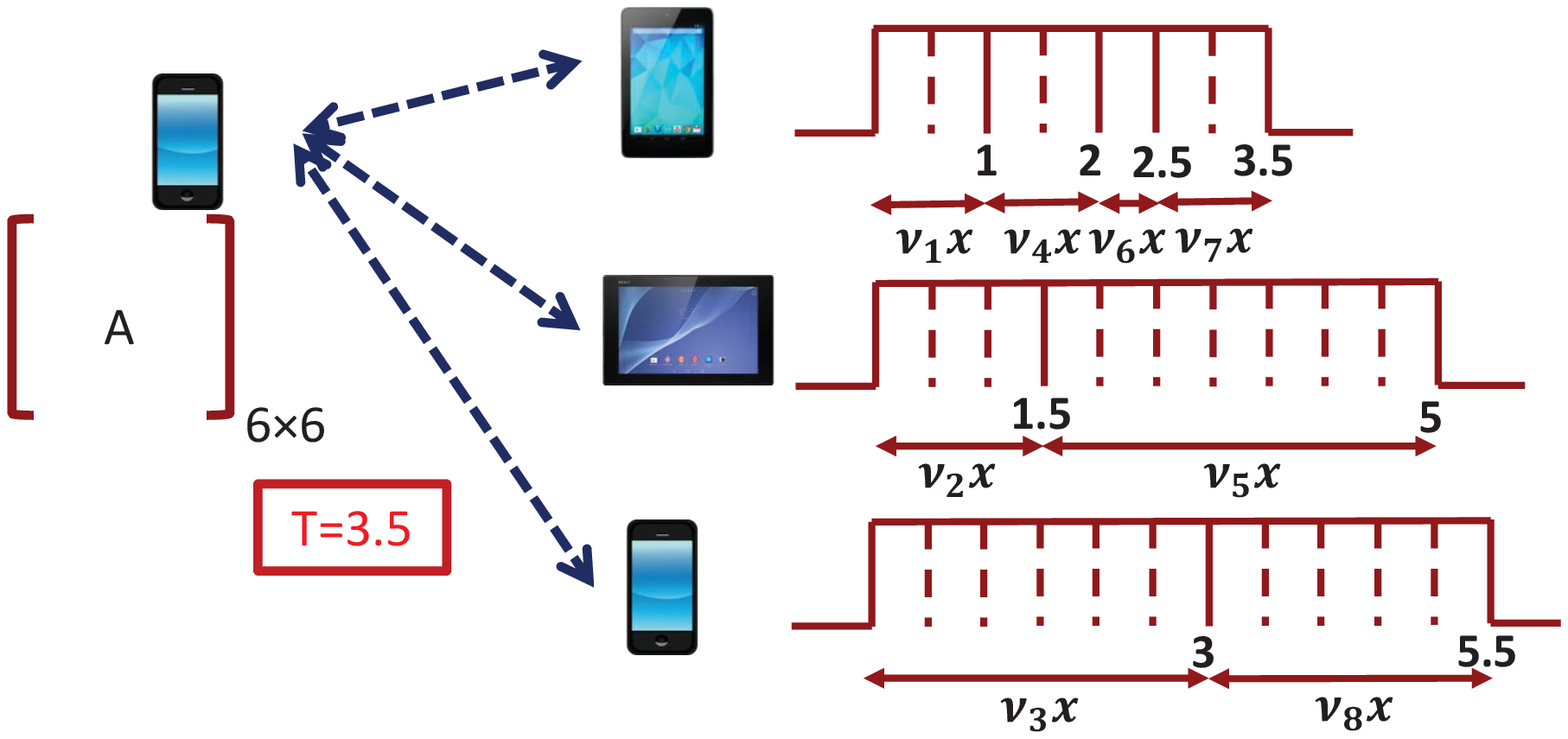}} }
\end{center}
\begin{center}
\caption{\label{fig:compare_rep_fountain} Performance of \CCP \ with respective to repetition codes with RR scheduling. }
\end{center}
\end{figure*}

As seen in Fig. \ref{fig:compare_rep_fountain}(a), RR scheduler sends $\rho_1$, $\rho_2$, and $\rho_3$ to helpers 1, 2, and 3, respectively at time $t = 0$. At time $t=1$, the computed packet $\rho_1 \mathbf{x}$ is received at the collector, and $\rho_4$, which is the next packet selected by RR scheduler, is transmitted to helper $1$. Similarly, at time $t=1.5$, $\rho_2 \mathbf{x}$ is received at the collector, and $\rho_5$ is transmitted to helper $2$. Similarly, the next packets are transmitted to helpers until the results for all packets are received at the collector, which is achieved at time $t=5$. As seen, the resources of helper $1$ is wasted while computing $\rho_3$, because those resources could have been used for computing a new packet. \CCP \ addresses this problem thanks to employing Fountain codes.

In particular, at time $t=0$, three Fountain coded packets of $\nu_1, \nu_2, \nu_3$ are created and transmitted to the three helpers, \ie $p_{1,1}=\nu_1, p_{2,1}=\nu_2, p_{3,1}=\nu_3$. At time $t=1$, a new coded packet of $\nu_4$ is created and transmitted as a second packet to helper $1$, \ie $p_{1,2}=\nu_4$.  This continues until $6$ computed coded packets (assuming that the overhead of Fountain codes, \ie $K$ is zero) are received at the collector, which is achieved at time $t=3.5$.
\hfill $\Box$
\end{example}

Example~\ref{ex:ex2} shows that the task completion delay is reduced from $5$ to $3.5$ when we use Fountain codes, which is significant. Section~\ref{sec:Simulation} shows extensive simulation results to support this illustrative example.

\subsection{Efficiency of \CCP} \label{sec:efficiency}
In this section, we characterize the efficiency of \CCP \ in the worst case scenario when per task runtimes follow the shifted exponential distribution. We call it the worst case efficiency, because we take into account per packet under-utilization $Tu_{n,i}$ in efficiency calculation, but  the fact that $Tu_{n,i}$ is monotonically decreasing, which is stated in Theorem \ref{theorem:CCP_vs_nonergodic}, is not used. 

\begin{theorem} \label{th:Theoretical_underutilized}
Assume that the runtime of each packet, \ie $\beta_{n,i}$, is a random variable according to an i.i.d shifted exponential distribution of
\begin{align}\label{eq:ShiftedExpDist}
  F_{\beta_{n,i}}(t) = Pr( \beta_{n,i} < t ) = 1 - e^{ - \mu_n (t - a_n)},
\end{align} with mean $a_n+1/\mu_n$ and shifted value of $a_n$. The expected value of the duration that helper $n$ is underutilized per packet is characterized as:
\begin{align} \label{eq:T_u_average_1}
    E[Tu_{n,i}]=
\begin{cases}
    &\hspace*{-3.5em}\frac{1}{(e\mu_n)}\Big(1-e^{(\mu_n RTT_n^{\text{\em data}})} \Big)+RTT_n^{\text{\em data}},\\
    &\text{\em if } RTT_n^{\text{\em data}} <\frac{1}{\mu_n}\\
    \frac{1}{(e\mu_n)},&\text{\em otherwise.}
\end{cases}
\end{align}
\end{theorem}
{\em Proof:}
The proof is provided in Appendix D. \hfill $\Box$

We define the efficiency of helper $n$ in the worst case as $\gamma_n = 1-{E[Tu_{n,i}]}/{E[\beta_{n,i}]}$. Note that $E[Tu_{n,i}]$ is the expected time that helper $n$ is underutilized per packet in the worst case, while $E[\beta_{n,i}]$ is the expected runtime duration, \ie the expected time that helper $n$ works per packet. Thus, ${E[Tu_{n,i}]}/{E[\beta_{n,i}]}$ becomes the under-utilization ratio of helper $n$ in the worst case, so $\gamma_n$ $=$ $1-{E[Tu_{n,i}]}$ $/$ ${E[\beta_{n,i}]}$ becomes the worst case efficiency. From (\ref{eq:T_u_average_1}) and replacing $E[\beta_{n,i}]$ with $a_n+1/\mu_n$, $\gamma_n$ is expressed as the following:
\begin{align}\label{eq:EfficincyAve}
  \gamma_n =
  \begin{cases}
    &\frac{1+a_n \mu_n -\mu_n RTT_n^{\text{data}}-1/e+\exp(\mu_n RTT_n^{\text{data}}-1)}{1+a_n\mu_n},\\
    & \quad \quad \quad \quad \quad \quad \quad \quad \quad \quad \text{if } RTT_n^{\text{data}}<1/\mu_n\\
    &\frac{e(1+a_n\mu_n)-1}{e(1+a_n\mu_n)}, \quad \quad \quad \quad \text{otherwise.}
\end{cases}
\end{align}
We show through simulations (in Section~\ref{sec:Simulation}) that, (i) $\gamma_n$ in (\ref{eq:EfficincyAve}) is larger than $99\%$, which is significant as (\ref{eq:EfficincyAve}) is the worst case efficiency,  and (ii) \CCP's efficiency is even larger than $\gamma_n$ as $\gamma_n$ in (\ref{eq:EfficincyAve}) is the efficiency in the worst case, where the under-utilization time period has the maximum value.

\section{\label{sec:Simulation} Performance Evaluation of \CCP}

In this section, we evaluate the performance of our algorithm; Coded Cooperative Computation Protocol (\CCP) via simulations and using real Android-based smartphones.

\subsection{Simulation Results}

We consider two scenarios: (i) Scenario 1, where the system resources for each helper vary over time. In this scenario, the runtime for computing each packet $p_{n,i}, \forall i$ at each helper $n$ is an i.i.d. shifted exponential random variable with shifted value $a_n$ and mean $a_n+1/\mu_n$, and
(ii) Scenario 2, where the runtime for computing packets in helper $n$ does not change over time, \ie $\beta_{n,i}=\beta_{n}, \forall i$, and $\beta_{n}, \forall n \in \Nset$ is a shifted exponential random variable with shifted value $a_n$ and mean $a_n+1/\mu_n$.

In our simulations, each simulated point is obtained by averaging over $200$ iterations for $N=100$ helpers. The transmission rate for sending each packet from the collector to each helper $n$ and from helper $n$ to the collector is a Poisson random variable with the average selected uniformly between $10$ Mbps and $20$ Mbps for each helper $n$.
The size of a transmitted packet $p_{n,i}$ is set to $B_x=8 R$ bits, where $R$ is the number of rows of matrix $A$, and it varies from $500$ to $20,000$ in our simulations. The sizes of a computed packet $p_{n,i} \mathbf{x}$ and an acknowledgement packet are set to $B_r=8$ bits and $B_{\text{ack}}=1$ bit, respectively. These are the parameters that are used for creating all plots unless otherwise is stated.

\begin{figure}[t!]
\centering
\subfigure[
Scenario 1]{ \scalebox{0.38}{\includegraphics{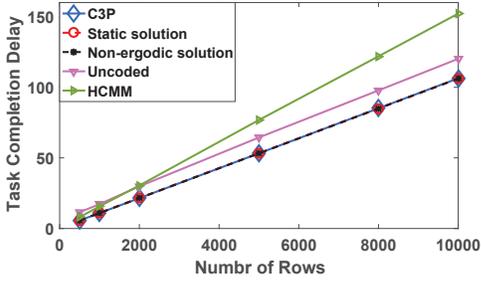}} }
\subfigure[
Scenario 2]{ \scalebox{.38}{\includegraphics{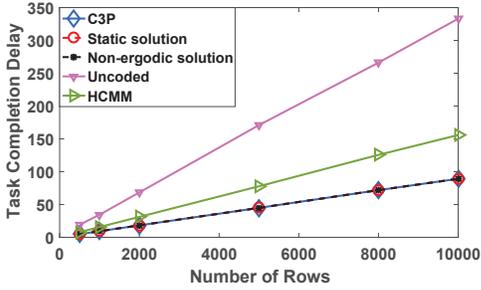}} }
\caption{
Task completion delay vs. number of rows/packets for (i) Scenario 1, and (ii) Scenario 2, where the runtime for computing one row by helper $n$ is selected from a shifted exponential distribution with $a_n=0.5, \forall n \in \Nset$ and $\mu_n$, which is selected uniformly from $\{1,2,4\}$.
}
\vspace{-15pt}
\label{fig:Runtime_vs_R_equalan}
\end{figure}

{\em Task Completion Delay vs. Number of Rows:}
We evaluate \CCP \  for Scenarios 1 and 2 and compare its task completion delay with: (i) Static solution, which is the task completion delay characterized in Section~\ref{sec:static} for both Scenarios 1 and 2. (ii) Non-ergodic solution, which is a realization of the non-ergodic problem characterized in Section \ref{sec:nonErgodic} by knowing $\beta_{n,i}$ a priori at the collector and setting $TTI_{n,i}$ as $\beta_{n,i}$. (iii) Uncoded: $r_n$ packets without coding are assigned to each helper $n$, and the collector waits to receive computed values from all helpers. The number of assigned packets to each helper $n$ is inversely proportional to the mean of $\beta_{n,i}$, \ie $r_n \propto \frac{1}{a_n+1/\mu_n}$. (iv) HCMM: Coded cooperative framework developed in \cite{HCMM} using block codes. We introduce $5\%$ coding overhead for \CCP, static, and non-ergodic solutions.

Fig. \ref{fig:Runtime_vs_R_equalan}(a) shows completion delay versus number of rows for Scenario 1, where the runtime for computing each packet by helper $n$, $\beta_{n,i}, \forall i$, is a shifted exponential random variable with shifted value of $a_n=0.5$ and mean of $a_n+1/\mu_n$, where $\mu_n$ is selected uniformly from $\{1,2,4\}$. As seen, \CCP \ performs close to the static and non-ergodic solutions. This shows the effectiveness of our proposed algorithm. In addition, \CCP \ performs better than the baselines. In particular, in average, $30\%$ and $24\%$ improvement is obtained by \CCP \ over HCMM and no coding, respectively. Fig. \ref{fig:Runtime_vs_R_equalan}(b) considers the same setup but for Scenario 2, where the runtime for computing $r_n$ packets by helper $n$ is $r_n \beta_{n}$, where $\beta_n$ is selected from a shifted exponential distribution with $a_n=0.5, \forall n \in \Nset$ and $\mu_n$, which is selected uniformly from $\{1,2,4\}$. As seen, for this scenario, \CCP \ performs close to the static and non-ergodic solutions. \CCP \ performs better than HCMM, and HCMM performs better than no coding. In particular, in average, $40\%$ and $69\%$ improvement is obtained by \CCP \ over HCMM and no coding, respectively.
Note that uncoded performs better than HCMM for Scenario 1, as HCMM is designed for Scenario 2, so it does not work well in Scenario 1. \CCP \ performs well in both scenarios.

\begin{figure}
[t!]
\centering
\subfigure[
Scenario 1]{ \scalebox{.38}{\includegraphics{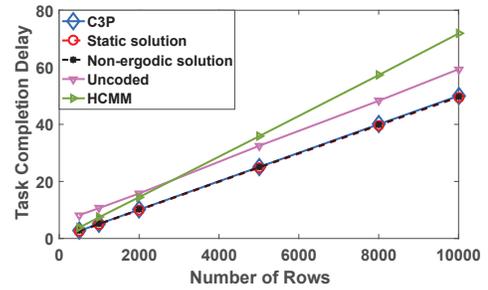}} }
\subfigure[
Scenario 2]{ \scalebox{.38}{\includegraphics{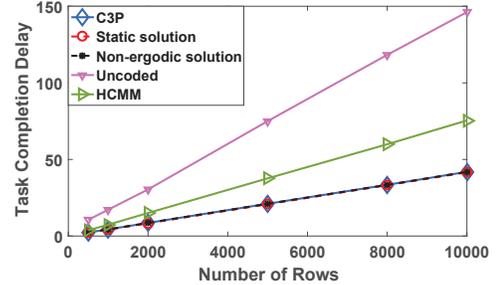}} }
\caption{
Task completion delay vs. number of rows/packets for (i) Scenario 1, and (ii) Scenario 2, where the runtime for computing one row by each helper $n$ is selected from a shifted exponential distribution with $\mu_n$, which is selected uniformly from $\{1,3,9\}$ for different helpers and $a_n=1/\mu_n, \forall n \in \Nset$
}
\vspace{-15pt}
\label{fig:Runtime_vs_R_nonequalan}
\end{figure}

Fig. \ref{fig:Runtime_vs_R_nonequalan} shows completion delay versus number of rows for both Scenarios 1 and 2, where the runtime for computing the rows by each helper $n$, is from a shifted exponential distribution with $\mu_n, n \in \Nset$ selected uniformly from $\{1,3,9\}$ and $a_n=1/\mu_n$ (different shifted values for different helpers).
As seen, \CCP \ performs close to static and non-ergodic solutions and much better than the baselines. In particular, for Scenario 1, more than $30\%$ and $15\%$ improvement is obtained by \CCP \ over HCMM and no coding, respectively. Also, for Scenario 2, in average, $42\%$ and $73\%$ improvement is obtained by \CCP \ over HCMM and no coding, respectively.

{\em Efficiency:}
We calculated the efficiency of helpers for different simulation setups and compared it with the theoretical efficiency obtained in (\ref{eq:EfficincyAve}) for Scenario 1. For all simulation setups, the average efficiency over all helpers was around $99\%$ and the theoretical efficiency was a little lower than the simulated efficiency. \Eg for $R=8000$ rows, where $\mu_n, n \in \Nset$ is selected uniformly from $\{1,3,9\}$ and $a_n=1/\mu_n$, the average of efficiency over all helpers is $99.7072\%$ and the average of theoretical efficiency is $99.4115\%$. This is expected as the theoretical efficiency is calculated for the worst case scenario.

We also calculate the efficiency of helpers for Scenario 2. For all simulation setups, the average efficiency over all helpers was around $99\%$, \eg for $R=8000$ rows, where $\mu_n, n \in \Nset$ is selected uniformly from $\{1,3,9\}$ and $a_n=1/\mu_n$, the average of efficiency over all helpers was $99.9267\%$. Note that the theoretical efficiency for Scenario 1 is $100\%$. The simulated efficiency is lower than the theoretical one, because the simulation underutilizes the helpers when transmitting the very first packet to each helper, \ie before the collector estimates the resources of helpers.

\begin{figure}
[t!]
\centering
{ \scalebox{.38}{\includegraphics{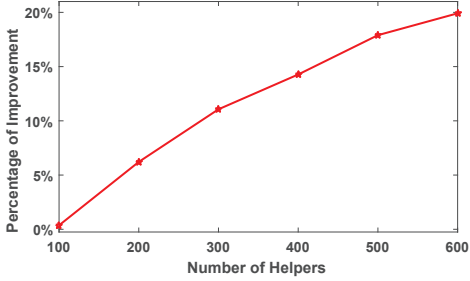}} }
\caption{
Percentage of improvement of \CCP \ over repetition codes with RR scheduling in terms of the task completion delay.
}
\vspace{-15pt}
\label{fig:Fountain_vs_Rep}
\end{figure}

{\em \CCP \ as Compared to Repetition Coding and Round Robin Scheduling:}
Fig. \ref{fig:Fountain_vs_Rep} shows the percentage of improvement of \CCP \ over repetition coding with RR scheduling in terms of task completion delay. The number of rows is selected as $R=2000$ with $5\%$ overhead for \CCP \ and the number of helpers varies from $N=100$ to $N=600$. The transmission rate for sending each packet from the collector to each helper $n$ and from helper $n$ to the collector is a Poisson random variable with the average selected uniformly between $0.1$ Mbps and $0.2$ Mbps for each helper $n$. The other parameters are the same as the parameters used in Fig. \ref{fig:Runtime_vs_R_equalan}(a). As seen, by increasing the number of helpers, more improvement is gained by \CCP \ compared to the repetition coding with RR scheduling. 

\subsection{Evaluation in a Testbed}
We implemented a testbed of a collector and multiple helpers using real mobile devices, specifically Android 6.0.1 based Nexus 6P and Nexus 5 smartphones. All the helpers are connected to the collector device using Wi-Fi Direct connections. We conducted our experiments using our testbed in a lab environment where several other Wi-Fi networks were operating in the background. We located all the devices in close proximity of each other (within a few meters distance).

We implemented both \CCP \ and repetition coding with RR scheduling in our testbed. The collector device would like to calculate matrix multiplication $\mathbf{y} = A\mathbf{x}$, where $A$ is a $1$K $\times$ $10$K matrix and $\mathbf{x}$ is a $10$K $\times$ $1$ vector. Matrix $A$ is divided into $20$ sub-matrices, each of which is a $50$ $\times$ $10$K matrix. A sub-task to be processed by a helper is the multiplication of a sub-matrix with vector $\mathbf{x}$. There is one collector device (Nexus 5) and varying number of helpers (Nexus 6P).

\begin{figure}
[t!]
\centering
\subfigure{ \scalebox{.32}{\includegraphics{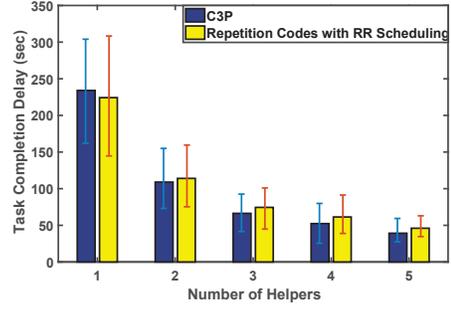}} }

\caption{
Task completion delay versus number of helpers.
}
\vspace{-10pt}
\label{fig:Runtime_vs_R_Android_1}
\end{figure}

Fig.~\ref{fig:Runtime_vs_R_Android_1} shows task completion delay versus number of helpers for both \CCP \ and repetition codes with RR scheduling. In this setup, each helper receives a sub-task, processes it, and  waits for a random amount of time (exponential random variable with mean $10$ seconds), which may arise due to other applications running at smartphones, and then sends the result back to the collector. As can be seen, the task completion delay reduces with increasing number of helpers in both algorithms. When there is one helper \CCP \ performs worse, which is expected. In particular, \CCP \ introduces coding overhead, and the number of helpers is very small to see the benefit of coding. On the other hand, when the number of helpers increases, we start seeing the benefit of coding. For example, when the number of helpers is $5$, \CCP \ improves $14\%$ over repetition codes with RR scheduling. This result confirms our simulation results in Fig. \ref{fig:Fountain_vs_Rep} in a testbed with real Android-based smartphones.

\begin{figure}
[t!]
\centering

\subfigure{ \scalebox{.32}{\includegraphics{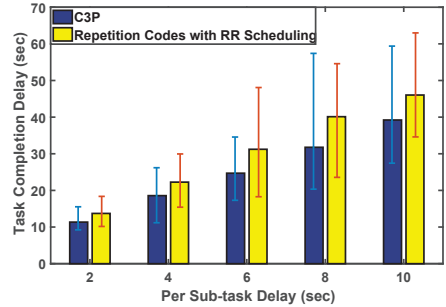}} }
\caption{
Task completion delay versus per sub-task delay.
}
\vspace{-15pt}
\label{fig:Runtime_vs_R_Android_2}
\end{figure}

Fig.~\ref{fig:Runtime_vs_R_Android_2} shows the task completion delay versus per sub-task random delays at helpers. There are $5$ helpers in this scenario. As can be seen, \CCP \ improves more over repetition codes with RR scheduling when delay increases, as it increases heterogeneity, and \CCP \ is designed to take into account heterogeneity.

\section{\label{sec:LitratureReview} Related Work}
Mobile cloud computing is a rapidly growing field with the goal of providing extensive computational resources to mobile devices as well as higher quality of experience \cite{L3, L4, L5}. The initial approach to mobile cloud computing has been to offload resource intensive tasks to remote clouds by exploiting Internet connectivity of mobile devices. This approach has received a lot of attention which led to extensive literature in the area \cite{L6, L7, L8, L9, L10}. The feasibility of computation offloading to remote cloud by mobile devices \cite{L11} as well as energy efficient computation offloading \cite{L12, L13} has been considered in the previous work. As compared to this line of work, our focus is on edge computing rather than remote clouds.

There is an increasing interest in edge computing by exploiting connectivity among mobile devices \cite{L14}. This approach suggests that if devices in close proximity are capable of processing tasks cooperatively, then local area computation groups could be formed and exploited for computation. 
Indeed, cooperative computation mechanisms by exploiting device-to-device connections of mobile devices in close proximity are developed in \cite{L14} and \cite{mobile_devices_3}.
A similar approach is considered in \cite{mobile_devices_computing_1} with particular focus on load balancing across workers. As compared to this line of work, we consider coded cooperative computation.

Coded cooperative computation is shown to provide higher reliability, smaller delay, and reduced communication cost in MapReduce framework \cite{MapReduce}, where computationally intensive tasks are offloaded to distributed server clusters \cite{CodedMapReduce}. In \cite{SpeedingUp} and \cite{ShortDot}, coded computation for matrix multiplication is considered, where matrix $A$ is divided into sub-matrices and each sub-matrix is sent from the master node (called collector in our work) to one of the worker nodes (called helpers in our work) for matrix multiplication with the assumption that the helpers are homogeneous. In \cite{SpeedingUp}, workload of the worker nodes is optimized such that the overall runtime is minimized. Fountain codes are employed in \cite{coded_Fountain} for coded computation, but for homogeneous resources. In \cite{HCMM}, the same problem is considered, but with the assumption that workers are heterogeneous in terms of their resources. Compared to this line of work, we develop \CCP, a practical algorithm that is (i) adaptive to the time-varying resources of helpers, and (ii) does not require any prior information about the computation capabilities of the helpers. As shown, our proposed method reduces the task completion delay significantly as compared to prior work.

\section{\label{sec:conclusion}Conclusion}
In this paper, we designed a Computation Control Protocol (\CCP), where heterogeneous edge devices with computation capabilities and energy resources are connected to each other. In \CCP, a collector device divides tasks into sub-tasks, offloads them to helpers by taking into account heterogeneous resources.  \CCP \ is (i) a dynamic algorithm that efficiently utilizes the potential of each helper, and (ii) adaptive to the time-varying resources at helpers. We analyzed the performance of \CCP \ in terms of task completion delay and efficiency. Simulation and experiment results in an Android testbed confirm that \CCP \ is efficient and reduces the completion delay significantly as compared to baselines. 

\bibliographystyle{acm}

\clearpage
\newpage

\section*{\label{sec:Calc_Exp_Collector} Appendix A: Calculating $E[\beta_{n,i}]$ at The Collector}
We define and use the parameters {\em residual time} $XTT_{n,i}$ and {\em round trip times of computed packets}; $RTT_{n,i}^{\text{data}}$ to estimate $E[\beta_{n,i}]$. Next, we will first show how we characterize $XTT_{n,i}$ and $RTT_{n,i}^{\text{data}}$, and then present how we estimate $E[\beta_{n,i}]$ using $XTT_{n,i}$ and $RTT_{n,i}^{\text{data}}$.

{\em Characterization of $XTT_{n,i}$.} The collector side has the knowledge of the transmission time $Tx_{n,i}$ of packet $p_{n,i}$, and time $Tr_{n,i}$ that the computed packet $p_{n,i} \mathbf{x}$ is received. Thus, the time between transmitting a packet and receiving its computed value, defined as $Tt_{n,i} = Tr_{n,i} - Tx_{n,i}$, can be calculated at the collector. On the other hand, to better utilize resources at helper $n$, the collector should offload a new packet before (or immediately after) receiving the computed value of the previous packet, \ie the following condition should be satisfied: $TTI_{n,i} \leq Tt_{n,i}$. Thus, we can write $TTI_{n,i} = Tt_{n,i} - XTT_{n,i+1}$, where $XTT_{n,i+1}$ is the residual time that the collector measures in our \CCP \ setup and is equal to:
\begin{align} \label{eq:XTT}
XTT_{n,i+1} = Tr_{n,i} - Tx_{n,i+1}.
\end{align}
\begin{figure}
[t!]
\centering
{ \scalebox{.25}{\includegraphics{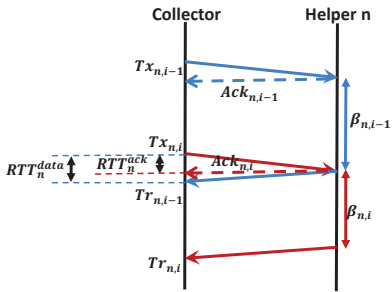}} }
\caption{
Demonstrating $RTT_n^{\text{data}}$ through an example.
}
\label{fig:RTT}
\end{figure}

{\em Characterization of $RTT_{n,i}^{\text{\em data}}$.}
We define $RTT_{n,i}^{\text{data}}$ as the round trip time (RTT) of packet $p_{n,i}$ sent to helper $n$. More precisely, $RTT_{n,i}^{\text{data}}$ is equal to transmission delay of packet $p_{n,i}$ from the collector to helper $n$ plus the transmission delay of the calculated packet $p_{n,i}\mathbf{x}$ from helper $n$ to the collector. Although $RTT_{n,i}^{\text{ data}}$ is round trip time, it can not be directly measured in the collector, as the collector only knows the time period between sending a packet and receiving the computed packet, which is equal to the sum of transmission and computing delay. Thus, in \CCP, we calculate $RTT_{n,i}^{\text{\em data}}$ using $RTT_{n,i}^{\text{ ack}}$, which is the time period between sending packet $p_{n,i}$ and receiving its ACK at the collector. Fig. \ref{fig:RTT} demonstrates the difference between $RTT_{n,i}^{\text{ data}}$ and $RTT_{n,i}^{\text{ ack}}$. Note that $RTT_{n,i}^{\text{ ack}}$ can be directly measured by employing ACKs. We can represent $RTT_{n,i}^{\text{ ack}}$ as
$RTT_{n,i}^{\text{ ack}} = {B_x}/{C_{n,i}^{\text{up}}}+{B_{\text{ack}}}/{C_{n,i}^{\text{down}}}$,

where $B_x$ is the size of the transmitted packet, $B_{\text{ack}}$ is the size of the ACK packet, and $C_{n,i}^{\text{up}}$ and $C_{n,i}^{\text{down}}$ are the uplink (from the collector to helper $n$) and downlink (from helper $n$ to the collector) transmission rates experienced by packet $p_{n,i}$ and its ACK.

Note that $RTT_{n,i}^{\text{ data}} $ is characterized as $RTT_{n,i}^{\text{ data}} = {B_x}/{C_{n,i}^{\text{up}}}+{B_r}/{C_{n,i}^{\text{down}}}$, where $B_r$ is the size of the computed packet; $p_{n,i}\mathbf{x}$. Assuming that uplink and downlink transmission rates are the same, which is likely in IoT setup, we can obtain $RTT_{n,i}^{\text{ data}}$ as:

\begin{align}\label{eq:XTTmin_2}
RTT_{n,i}^{\text{ data}} = \frac{B_x+B_r}{B_x+B_{\text{ack}}}RTT_{n,i}^{\text{ ack}}.
\end{align}
As we discussed earlier $RTT_{n,i}^{\text{ ack}}$ can be directly measured by the collector and used in (\ref{eq:XTTmin_2}) to determine $RTT_{n,i}^{\text{ data}}$. Next, we characterize the average value of data round trip time of helper $n$, \ie $RTT_{n}^{\text{ data}}$ as exponential weighted moving averages of per packet round trip time, $RTT_{n,i}^{\text{ data}}$:
\begin{align}\label{eq:XTTmin_3}
 RTT_{n}^{\text{ data}} = \alpha RTT_{n,i}^{\text{ data}}  + (1-\alpha) RTT_{n}^{\text{ data}},
\end{align} where $\alpha$ is a weight satisfying $0 < \alpha < 1$. 
Now that we characterized $XTT_{n,i}$ and $RTT_{n}^{\text{ data}}$ and discussed how we can measure these parameters at the collector, we explain how to use these parameters to calculate $E[\beta_{n,i}]$ at the collector.

{\em Calculation of $E[\beta_{n,i}]$.}
We formulate $E[\beta_{n,i}]$ as follows:
\begin{align} \label{eq:average_beta}
E[\beta_{n,i}] \approx \frac{\sum_{j=1}^{m_n} \beta_{n,i}}{m_n} = \frac{Tc_{n,i} - Tu_{n}}{m_n},
\end{align} where $Tc_{n,i}$ is the time that helper $n$ finishes computing $p_{n,i}\mathbf{x}$, $Tu_{n}$ is the estimate made at the collector about the total (cumulative) time that helper $n$ is underutilized, and $m_n$ is the number of packets that helper $n$ processed until (and including) packet $p_{n,i}$. Since $Tc_{n,i}$ is the time instance that helper $n$ finishes computing packet $p_{n,i} \mathbf{x}$, and $Tu_n$ is the cumulative time that helper $n$ is underutilized, their difference gives us the total time that helper $n$ has been busy since the starting time. Total busy time of helper $n$, \ie $Tc_{n,i} - Tu_{n}$ is normalized by the total number of processed packets $m_n$ to determine $E[\beta_{n,i}]$. Next, we characterize $Tc_{n,i}$ and $Tu_{n}$ in terms of $XTT_{n,i}$ and $RTT_{n}^{\text{ data}}$.

The collector estimates $Tc_{n,i}$ as follows
\begin{align}\label{eq:Tci}
Tc_{n,i} \approx Tr_{n,i} - \frac{B_r}{B_x + B_r} RTT_{n}^{\text{data}},
\end{align} where $Tr_{n,i}$ is the time that the computed packet $p_{n,i} \mathbf{x}$ is received by the collector from helper $n$, so it is known by the collector. $\frac{B_r}{B_x + B_r} RTT_n^{\text{data}}$ is the backward trip time estimated by the collector using (\ref{eq:XTTmin_3}) and packet sizes.

The next step is to characterize $Tu_n$, which is the estimate made at the collector about the total (cumulative) time that helper $n$ is underutilized. $Tu_n$ is the sum of all per packet under-utilization times, shown by $Tu_{n,i}$. In particular, $Tu_{n,i}$ is defined as the time period that the helper is idle between computing packet $p_{n,i-1} \mathbf{x}$ and $p_{n,i} \mathbf{x}$. In order to calculate $Tu_{n,i}$, we should determine the state of the system, \ie if the system is in the ideal, underutilized, or congested case, Fig. \ref{fig:CPP_Scenarios}. As shown in Fig. \ref{fig:CPP_Scenarios}, $XTT_{n,i} \geq RTT_n^{\text{data}}$ in the ideal and congested cases. Note that these cases occur when packet $p_{n,i}$ is received at the helper meanwhile the helper is computing $p_{n,i-1} \mathbf{x}$ or right after the helper has computed $p_{n,i-1} \mathbf{x}$. In this setup, since there is no under-utilization, $Tu_{n,i}$ is equal to $0$. On the other hand, $XTT_{n,i} < RTT_n^{\text{data}}$ in the underutilized case scenario. As seen in Fig. \ref{fig:CPP_Scenarios}(b), underutilized case occurs when packet $p_{n,i}$ is received at the helper after a while that the helper has finished computing packet $p_{n,i-1} \mathbf{x}$. In this setup, $RTT_n^{\text{data}} - XTT_{n,i}$ is the approximate duration that helper $n$ is idle before calculating $p_{n,i} \mathbf{x}$, \ie $Tu_{n,i} \approx RTT_n^{\text{data}} - XTT_{n,i}$. Therefore, $Tu_n$ is updated after $p_{n,i}\mathbf{x}$ is received by the collector as the following:
\begin{align}\label{eq:Tui}
Tu_n \approx Tu_n + \max\{0, RTT_n^{\text{data}} - XTT_{n,i}\}.
\end{align}

As seen, $E[\beta_{n,i}]$ can be calculated from the parameters that are known by the collector. The process of calculating $E[\beta_{n,i}]$ by the collector is summarized in Algorithm \ref{al:calc_E}.

\begin{algorithm}[t!]
\caption{Calculating $E[\beta_{n,i}]$ by the collector}
\label{al:calc_E}
\begin{algorithmic}[1]
\STATE Initialize: $Tx_{n,1}=0$, $RTT_n^{\text{data}}=0$, $\forall n \in \Nset$.

\IF {An ACK for successful transmission of packet $p_{n,i}$ is received from helper $n$}
	\STATE Update $RTT_n^{\text{data}}$ according to (\ref{eq:XTTmin_3}). \label{al:XTT_min_3rd}
 \ENDIF
 \IF {Calculated packet $p_{n,i} \mathbf{x}$ and the corresponding computation ACK is received}
	\IF {$i==1$} \label{al:XTT_i_1}
		\STATE $Tu_n=0$.
	\ELSE
        \STATE $XTT_{n,i}=Tr_{n,i-1}-Tx_{n,i}$. \label{al:XTT_i_2}
        \STATE Update $Tu_n$ according to (\ref{eq:Tui}).
    \ENDIF
\ENDIF
 \STATE Calculate $E[\beta_{n,i}]$ from (\ref{eq:average_beta}).
 
\end{algorithmic}
\end{algorithm}

\section*{\label{sec:Ape_underutilized_conditions} Appendix B: Proof of Lemma \ref{th:underutilized_conditions}}
To prove Lemma \ref{th:underutilized_conditions}, first we characterise $Tu_{n,i}$ and then find the closed form conditions for $Tu_{n,i}>0$.

\subsection{Characterising $Tu_{n,i}$}
According to (\ref{eq:TTI_in}), in \CCP \ the packets are transmitted from the collector to helper $n$ with the time interval equal to $\min ( Tr_{n,i} - Tx_{n,i}, E[ \beta_{n,i} ] )$. We first provide the queuing model for the case of $TTI_{n,i}$ equal to $E[ \beta_{n,i} ]$ and then show that the queueing model for \CCP \ is the same as this queue with the only difference that the idle time in \CCP \ is smaller than the idle time for the case with $TTI_{n,i}$ equal to $E[ \beta_{n,i} ]$.

{\em Queueing model for $TTI_{n,i}$ equal to $E[ \beta_{n,i} ]$.} The system of the collector and helper $n$ for the case with $TTI_{n,i}$ equal to $E[ \beta_{n,i} ]$ can be modeled as a queue, where each packet $p_{n,i}$ is arrived at helper $n$ with the arrival rate of $1$ packet per $E[\beta_{n,i}]$ and processed with the service time of $\beta_{n,i}$. In the steady state case, (\ie the case that the queue is empty at the time packet $p_{n,i}$ is received at the helper), if the service time is larger than the arrival time, \ie $\beta_{n,i}>E[\beta_{n,i}]$, the next received packet of $p_{n,i+1}$ will be queued at the helper for the time period equal to the difference between the service time and the arrival time, \ie $\beta_{n,i}-E[\beta_{n,i}]$. On the other hand, if the service time is smaller than the arrival time, \ie $\beta_{n,i}<E[\beta_{n,i}]$, processing of the received packet $p_{n,i+1}$ will be delayed for the time period equal to the difference between the arrival time and the service time, \ie $E[\beta_{n,i}]-\beta_{n,i}$, after computing the previous packet of $p_{n,i}$. This is the idle (underutilized) time period of the queue. Now let us consider the general case where the queue is not empty when packet $p_{n,i+1}$ is received at helper $n$ with the queueing delay of $Tq_{n,i}$, \ie it takes $Tq_{n,i}$ for helper $n$ to start computing the last packet in its queue. In this case, if $\beta_{n,i}<E[\beta_{n,i}]$, then the underutilized time period between computing packet $p_{n,i}$ and packet $p_{n,i+1}$ at the helper is equal to $max(0,E[\beta_{n,i}]-\beta_{n,i}-Tq_{n,i})$ and thus $Tu_{n,i}$ is characterized as
\begin{align}\label{Tu_general}
  Tu_{n,i} = \max\big(\max(0,E[\beta_{n,i}]-\beta_{n,i})-Tq_{n,i},0\big).
\end{align}We will formulate $Tq_{n,i}$ later in this section, but before that let us formulate $Tu_{n,i}$ for \CCP.

{\em Formulating $Tu_{n,i}$ for \CCP.} The difference between \CCP \ and the case when $TTI_{n,i}$ is equal to $E[ \beta_{n,i} ]$, is that in \CCP \ the idle time is reduced. In particular, if the collector notices that the helper is idle (by receiving the computed packet $p_{n,i}\mathbf{x}$ before sending packet $p_{n,i+1}$), it reduces $TTI_{n,i}$, the time interval between sending packet $p_{n,i}$ and $p_{n,i+1}$, to $Tr_{n,i} - Tx_{n,i}$. In this case, from Fig. \ref{fig:CPP_Scenarios}(b), the parameter $XTT_{n,i}$ becomes zero, which results in reduced underutilized time of $Tu_{n,i}$ to $RTT_n^{\text{data}}$. Therefore, $Tu_{n,i}$ for \CCP \ is equal to:
\begin{align} \label{eq:under_utilization}
Tu_{n,i} = \min\Big(&\max\big(\max(0,E[\beta_{n,i}]-\beta_{n,i})-Tq_{n,i},0\big), \nonumber\\
&RTT_n^{\text{data}}\Big).
\end{align}

{\em Formulating $Tq_{n,i}$.} The queueing delay $Tq_{n,i}$ is defined as the period that packet $p_{n,i}$ should wait in the queue to be computed by helper $n$. We consider two cases to calculate $Tq_{n,i}$: (i) $\beta_{n,i-1}>E[\beta_{n,i}]$: this is the congested case scenario, where $p_{n,i}$ is received at the helper while the helper is busy computing the previously received packets. Therefore, packet $p_{n,i}$ should be queued at the helper queue and its queueing delay is equal to the summation of $\beta_{n,i-1}-E[\beta_{n,i}]$ and the queueing delay for computing its previous packet $p_{n,i-1}$, which is equal to $Tq_{n,i-1}$ and thus $Tq_{n,i} = \beta_{n,i-1}-E[\beta_{n,i}]+Tq_{n,i-1}$. (ii) $\beta_{n,i-1}<E[\beta_{n,i}]$: this is the underutilized case scenario if there is no packet in the queue when packet $p_{n,i}$ is received at the collector. In this case, the helper will be idle for the time period of $E[\beta_{n,i}]-\beta_{n,i-1}$ after it computes packet $p_{n,i-1}$ until it receives packet $p_{n,i}$ and starts computing it. However, if there are packets in the queue at the time packet $p_{n,i}$ is received at helper $n$, then two cases may occur: (a) $Tq_{n,i-1}-(E[\beta_{n,i}]-\beta_{n,i-1})>0$: in this case, packet $p_{n,i}$ still should wait in the queue but its queueing delay is reduced compared to the queueing delay for packet $p_{n,i-1}$ by $E[\beta_{n,i}]-\beta_{n,i-1}$. (b) $Tq_{n,i-1}-(E[\beta_{n,i}]-\beta_{n,i-1})<0$: in this case, the queueing delay for packet $p_{n,i}$ is zero and $p_{n,i}$ will be computed by helper $n$ as soon as it is received at the helper. The reason is that helper $n$ finishes computing the previous packet $p_{n,i-1}$ earlier than packet $p_{n,i}$ is received at the helper and remains idle for the period of $(E[\beta_{n,i}]-\beta_{n,i-1})-Tq_{n,i-1}$ before it starts computing packet $p_{n,i}$. By considering all these cases, $Tq_{n,i}$ can be formulated as:
\begin{align} \label{eq:queueing_elay}
Tq_{n,i} = \max(\beta_{n,i-1}-E[\beta_{n,i}]+Tq_{n,i-1},0).
\end{align} 

\subsection{Finding The Closed Form Conditions For $Tu_{n,i}>0$}
From (\ref{eq:under_utilization}), for $Tu_{n,i+1}$ to be positive, \ie $\max(\max(0,E[\beta_{n,i}]-\beta_{n,i})-Tq_{n,i},0)>0$, the following condition should be satisfied:
\begin{align} \label{eq:cond_k=1_1}
 &\max(0,E[\beta_{n,i}]-\beta_{n,i})-Tq_{n,i} > 0 \\
 & \Leftrightarrow \max(0,E[\beta_{n,i}]-\beta_{n,i}) > Tq_{n,i}\\
 & \Leftrightarrow E[\beta_{n,i}]-\beta_{n,i} > Tq_{n,i} \label{eq:cond_induc_k=1}
\end{align}
By replacing $Tq_{n,i}$ from (\ref{eq:queueing_elay}), we have:
\begin{align}
E[\beta_{n,i}]-\beta_{n,i} > \max(\beta_{n,i-1}-E[\beta_{n,i}]+Tq_{n,i-1},0)
\end{align}
\begin{numcases}{\Leftrightarrow}
&\hspace*{-2em}$E[\beta_{n,i}]-\beta_{n,i} > 0$\\
&\hspace*{-2em}$E[\beta_{n,i}]-\beta_{n,i} > \beta_{n,i-1}-E[\beta_{n,i}]+Tq_{n,i-1}$
\end{numcases}
\begin{numcases}{\Leftrightarrow}
&\hspace*{-2em}$E[\beta_{n,i}] > \beta_{n,i}$ \label{eq:cond_k=1}\\
&\hspace*{-2em}$2E[\beta_{n,i}]-\beta_{n,i}-\beta_{n,i-1} > Tq_{n,i-1}$ \label{eq:cond_k=2_1}
\end{numcases}
(\ref{eq:cond_k=1}) generates the first condition of Lemma \ref{th:underutilized_conditions}, \ie $k=1$. By replacing $Tq_{n,i-1}$ in (\ref{eq:cond_k=2_1}), we have:
\begin{align}
  2E[\beta_{n,i}]-\beta_{n,i}-\beta_{n,i-1} > \max(&\beta_{n,i-2}-E[\beta_{n,i}]+Tq_{n,i-2}, \nonumber\\
  &0)
\end{align}

\begin{numcases}{\Leftrightarrow}
 &\hspace*{-2em}$2E[\beta_{n,i}] > \beta_{n,i}+\beta_{n,i-1}$ \label{eq:cond_k=2}\\
 &\hspace*{-2em}$3E[\beta_{n,i}]-\beta_{n,i}-\beta_{n,i-1}-\beta_{n,i-2} > Tq_{n,i-2}$\label{eq:cond_others}
\end{numcases}
(\ref{eq:cond_k=2}) generates the second condition of Lemma \ref{th:underutilized_conditions}, \ie $k=2$. Intuitively, we can prove all other conditions of Lemma \ref{th:underutilized_conditions} by replacing $Tq_{n,i-2}$ in (\ref{eq:cond_others}). In the following, we give a formal proof using the proof by induction.

First we prove by induction that $(kE[\beta_{n,i}]-\sum_{j=i-k+1}^{i} \beta_{n,j}) > Tq_{n,i-k+1}$ is satisfied for $\forall k=1,2,\ldots,i-1$ when $Tu_{n,i}>0$; We already showed in (\ref{eq:cond_induc_k=1}) that this is true for $k=1$. If this inequality is true for $k=m$, \ie $(mE[\beta_{n,i}]-\sum_{j=i+1-m}^{i} \beta_{n,j}) > Tq_{n,i-m+1}$, then by replacing $Tq_{n,i-m+1}$ with its equivalent from (\ref{eq:queueing_elay}), we have:
\begin{align}\label{eq:cond_induct}
  & (mE[\beta_{n,i}]-\sum_{j=i+1-m}^{i} \beta_{n,j}) > \\
  & \quad \quad \quad \quad \max(\beta_{n,i-m}-E[\beta_{n,i}]+Tq_{n,i-m},0) \nonumber\\
  \Rightarrow & (mE[\beta_{n,i}]-\sum_{j=i+1-m}^{i} \beta_{n,j}) > \\
  & \quad \quad \quad \quad (\beta_{n,i-m}-E[\beta_{n,i}]+Tq_{n,i-m}) \nonumber\\
  \Rightarrow & ((m+1)E[\beta_{n,i}]-\beta_{n,i-m})-\sum_{j=i+1-m}^{i} \beta_{n,j}>Tq_{n,i-m}\\
  \Rightarrow & ((m+1)E[\beta_{n,i}]-\sum_{j=i-m}^{i} \beta_{n,j}>Tq_{n,i-m},
\end{align}
and thus the inequality is true for $k=m+1$. This proves $(kE[\beta_{n,i}]-\sum_{j=i+1-k}^{i} \beta_{n,j}) > Tq_{n,i-k+1}, \forall k=1,2,\ldots,i$ from which we conclude $(kE[\beta_{n,i}] > \sum_{j=i+1-k}^{i} \beta_{n,j}), \forall k=1,2,\ldots,i$ as $Tq_{n,i-k}$ is positive. Therefore, we proved that the $i$ conditions of $(kE[\beta_{n,i}] > \sum_{j=i+1-k}^{i} \beta_{n,j}), \forall k=1,2,\ldots,i$ is necessary for $Tu_{n,i}>0$. 

Now, we prove the sufficiency of conditions in Lemma \ref{th:underutilized_conditions}; \ie if $(kE[\beta_{n,i}] > \sum_{j=i+1-k}^{i} \beta_{n,j}), \forall k=1,2,\ldots,i$, then $Tu_{n,i+1}>0$ or equivalently $(E[\beta_{n,i}]-\beta_{n,i}) > Tq_{n,i}$. First we prove by induction that $(i-l)E[\beta_{n,i}]-\sum_{j=l+1}^{i} \beta_{n,j} > Tq_{n,l+1}$ is satisfied for $\forall l=0,1,\ldots,i-1$, when $(kE[\beta_{n,i}] > \sum_{j=i+1-k}^{i} \beta_{n,j}), \forall k=1,2,\ldots,i$; For $l=0$, we just need to make $k$ equal to $i$ in $(kE[\beta_{n,i}] > \sum_{j=i+1-k}^{i} \beta_{n,j})$, as the queueing delay for the first received packet at helper $n$, $Tq_{n,1}$, is zero. Now we assume that $(i-l)E[\beta_{n,i}]-\sum_{j=l+1}^{i} \beta_{n,j} > Tq_{n,l+1}$ is satisfied for $l=m$:
\begin{align}
  &(i-m)E[\beta_{n,i}]-\sum_{j=m+1}^{i} \beta_{n,j} > Tq_{n,m+1} \\
  \Rightarrow & (i-m-1)E[\beta_{n,i}]-\sum_{j=m+2}^{i} \beta_{n,j} > \nonumber\\
  & \quad \quad \quad \quad \beta_{n,m+1}-E[\beta_{n,i}]+Tq_{n,m+1}
\end{align}
On the other hand, by replacing $k=i-m-1$ in $(kE[\beta_{n,i}] > \sum_{j=i+1-k}^{i} \beta_{n,j})$, we have $(i-m-1)E[\beta_{n,i}]-\sum_{j=m+2}^{i} \beta_{n,j}>0$, and thus we have:
\begin{align}
 &(i-m-1)E[\beta_{n,i}]-\sum_{j=m+2}^{i} \beta_{n,j} \nonumber\\
 & \quad \quad \quad \quad > \max(\beta_{n,m+1}-E[\beta_{n,i}]+Tq_{n,m+1},0)\\
 & \quad \quad \quad \quad = Tq_{n,m+2}.
\end{align}
The above inequality shows that $(i-l)E[\beta_{n,i}]-\sum_{j=l+1}^{i} \beta_{n,j} > Tq_{n,l+1}$ is satisfied for $l=m+1$. Therefore, from proof by induction, $(i-l)E[\beta_{n,i}]-\sum_{j=l+1}^{i} \beta_{n,j} > Tq_{n,l+1}$ is satisfied for $\forall l=0,1,\ldots,i-1$. By replacing $l=i-1$ in $(i-l)E[\beta_{n,i}]-\sum_{j=l+1}^{i} \beta_{n,j} > Tq_{n,l+1}$, we have $(E[\beta_{n,i}]-\beta_{n,i}) > Tq_{n,i}$ or equivalently $Tu_{n,i+1}>0$. This proves the sufficiency of conditions in Lemma \ref{th:underutilized_conditions}.

This concludes the proof. 
\section*{Appendix C: Proof of Theorem \ref{theorem:CCP_vs_static}}
First we show that the queueing delay of $Tq_{n,i}$ is small and then use this property to prove Theorem \ref{theorem:CCP_vs_static}.

According to (\ref{eq:queueing_elay}), the queueing delay $Tq_{n,i}$, which is the delay for packet $p_{n,i}$ to be computed by helper $n$, is equal to the sum of $(E[\beta_{n,i}]-\beta_{n,i-1})$ and $Tq_{n,i-1}$, if this sum is positive, otherwise it is qual to zero. If we look at this equation more closely, we observe that we can reformulate $Tq_{n,i}$ as $\sum_{j=i'}^{i-1}(E[\beta_{n,i}]-\beta_{n,j})$, where $i'<i-1$ corresponds to the last time that helper $n$ has been seen as underutilized, \ie $i'$ is the largest $j<i-1$, for which $Tq_{n,j-1}=0$. Therefore, the average of $Tq_{n,i}$ is equal to $E[Tq_{n,i}]=$ $E[\sum_{j=i'}^{i-1}(E[\beta_{n,i}]-\beta_{n,j})]=0$. Therefore, in average, the queueing delay of \CCP \ is zero. Next, we use this property to prove Theorem \ref{theorem:CCP_vs_static}.

\CCP \ sends packets to each helper $n$ with the time interval less than or equal to $E[\beta_{n,i}]$ according to (\ref{eq:TTI_in}). Since the queueing delay of \CCP \ is small, this time interval will result in the delay of $D_n^{\text{\CCP}}$ less than or equal to $r_n^{\text{\CCP}} E[\beta_{n,i}]$ for calculating $r_n^{\text{\CCP}}$ packets by helper $n$. On the other hand, the collector stops sending packets to helpers once it receives $R+K$ packets collectively from all helpers. Again, since the queueing delay of \CCP \ is small, \ie the period of time packets wait in the queue of helper $n$ to be computed is small, at the time that $R+K$ packets are collected at the collector, there might be only small number of packets waiting at the queue of each helper and thus $\sum_{n=1}^{N} r_n^{\text{\CCP}}\simeq R+K$. In addition, with small queueing delay, all helpers will finish their assigned tasks approximately at the same time; this results in (\ref{eq:TCCP}) and (\ref{eq:rnCCP}).

This concludes the proof. \hfill $\Box$ 

\section*{\label{sec:AppendixA} Appendix D: Proof of Theorem \ref{th:Theoretical_underutilized}}
With the assumption that $\beta_{n,i-1}$ is from a shifted exponential distribution with shifted value of $a_n$ and mean of $a_n+1/\mu_n$, $E[\beta_{n,i}]$ in (\ref{eq:under_utilization}) can be replaced with $a_n+1/\mu_n$. In addition, $Tq_{n,i}$ is equal to zero as we consider the worst case scenario. Therefore, we have:
\begin{align} \label{eq:T_u_33}
Tu_{n,i}=
\begin{cases}
    &\hspace*{-1.75em}RTT_n^{\text{data}}, \\
    &\text{if } a_n<\beta_{n,i-1}<a_n+\frac{1}{\mu_n}-RTT_n^{\text{data}}\\
    &\hspace*{-1.75em}a_n+\frac{1}{\mu_n}-\beta_{n,i-1}, \\
    &\text{if } a_n+\frac{1}{\mu_n}-RTT_n^{\text{data}}<\beta_{n,i-1}<a_n+\frac{1}{\mu_n} \\
    0, &\text{otherwise,}
\end{cases}
\end{align}
where, the random variable $\beta_{n,i-1}$ is always greater than its shifted value, $a_n$, \ie the condition $\beta_{n,i-1}>a_n$ is always satisfied. From (\ref{eq:T_u_33}), the value of variable $Tu_{n,i}$ changes depending on the value of $RTT_n^{\text{data}}$ and the value of the distribution parameter $1/\mu_n$. We decompose (\ref{eq:T_u_33}) as follows:
\begin{align} \label{eq:T_u}
    Tu_{n,i}=
\begin{cases}
    \tilde{T}_{n,i},              & \text{if } RTT_n^{\text{data}}<1/\mu_n\\
    \overline{T}_{n,i}, & \text{otherwise,}
\end{cases}
\end{align}
where,
\begin{align} \label{eq:T_u_11}
    \tilde{T}_{n,i}=
\begin{cases}
    &\hspace*{-1.75em}RTT_n^{\text{data}},\\
    &\text{if } a_n<\beta_{n,i-1}<a_n+\frac{1}{\mu_n}-RTT_n^{\text{data}}\\
    &\hspace*{-1.75em}a_n+\frac{1}{\mu_n}-\beta_{n,i-1},\\
    &\text{if } a_n+\frac{1}{\mu_n}-RTT_n^{\text{data}}<\beta_{n,i-1}<a_n+\frac{1}{\mu_n}\\
    0, &\text{otherwise,}
\end{cases}
\end{align}
and,
\begin{align} \label{eq:T_u_2}
    \overline{T}_{n,i}=
\begin{cases}
    a_n+1/\mu_n-\beta_{n,i-1},              & \text{if } a_n \leq \beta_{n,i-1} \leq a_n+1/\mu_n\\
    0, & \text{otherwise}
\end{cases}
\end{align}
To prove Theorem \ref{th:Theoretical_underutilized}, we find the expected values of $\tilde{T}_{n,i}$ and $\overline{T}_{n,i}$.
From (\ref{eq:T_u_11}), the average of $\tilde{T}_{n,i}$ is calculated as:
\begin{align}\label{eq:TUAve_1_derivation}
  E[\tilde{T}_{n,i}] & = \int f_{\beta_{n,i-1}}(t) \tilde{T}_{n,i} dt\nonumber\\
  &\hspace*{-2em} = \int_{a_n}^{a_n+1/\mu_n-RTT_n^{\text{data}}} \mu_n \exp(-\mu_n(t-a_n)) RTT_n^{\text{data}} dt \nonumber \\
  &\hspace*{-2em} + \int_{a_n+\frac{1}{\mu_n}-RTT_n^{\text{data}}}^{a_n+\frac{1}{\mu_n}} \mu_n \exp(-\mu_n(t-a_n))(a_n+\frac{1}{\mu_n}-t) dt \nonumber\\
  &\hspace*{-2em} =RTT_n^{\text{data}}+1/\mu_n(\exp(-1)-\exp(\mu_n RTT_n^{\text{data}}-1)).
\end{align}
Similarly, from (\ref{eq:T_u_2}), we can calculate the average of $\overline{T}_{n,i}$:
\begin{align}\label{eq:TUAve_2_derivation}
  E[\overline{T}_{n,i}] & = \int f_{\beta_{i-1}}(t) \overline{T}_{n,i} dt\nonumber\\
  & \hspace*{-1em}= \int_{a_n}^{a_n+1/\mu_n} \mu_n \exp(-\mu_n(t-a_n)) (a_n+1/\mu_n-t) dt \nonumber\\
  & \hspace*{-1em}= 1/\mu_n \exp(-1).
\end{align}
By replacing the obtained expected values of $\tilde{T}_{n,i}$ and $\overline{T}_{n,i}$ in (\ref{eq:T_u}), we have:
\begin{align} \label{eq:T_u_average}
    E[Tu_{n,i}]=
\begin{cases}
    &\hspace*{-4em}RTT_n^{\text{data}}+\frac{1}{\mu_n}(e^{-1}-\exp(\mu_n RTT_n^{\text{data}}-1)), \\
    & \text{if } RTT_n^{\text{data}}<\frac{1}{\mu_n}\\
    \frac{1}{\mu_n} e^{-1}, & \text{otherwise.}
\end{cases}
\end{align}
This concludes the proof.

\end{document}